\documentclass[sigconf]{aamas} 

\usepackage{censor}
\usepackage{balance} 

\def\bl{}
\def\censorcolor{white}
\let\svcensorrule\censorrule
\renewcommand\censorrule[1]{%
\textcolor{\censorcolor}{\svcensorrule{#1}}}

\setcopyright{ifaamas}
\acmConference[AAMAS '23]{\bl{Proc.\@ of the 22nd International Conference
on Autonomous Agents and Multiagent Systems (AAMAS 2023)}}{\bl{May 29 -- June 2, 2023}}
{\bl{London, United Kingdom}}{\bl{A.~Ricci, W.~Yeoh, N.~Agmon, B.~An (eds.)}}
\copyrightyear{2023}
\acmYear{2023}
\acmDOI{}
\acmPrice{}
\acmISBN{}

\begin{abstract}
    Measures of allocation optimality differ significantly when distributing standard tradable goods in peaceful times and scarce resources in crises. While realistic markets offer asymptotic efficiency, they may not necessarily guarantee fair allocation desirable when distributing the critical resources. To achieve fairness, mechanisms often rely on a central authority, which may act inefficiently in times of need when swiftness and good organization are crucial. In this work, we study a hybrid trading system called {\em Crisdis}, introduced by Jedli\v{c}kov\'{a} et al., which combines fair allocation of buying rights with a market -- leveraging the best of both worlds. A~frustration of a buyer in Crisdis is defined as a difference between the amount of goods they are entitled to according to the assigned buying rights and the amount of goods they are able to acquire by trading. We define a Price of Anarchy (PoA) in this system as a conceptual analogue of the original definition in the context of frustration.
    Our main contribution is a study of PoA in realistic complex double-sided market mechanisms for Crisdis. The performed empirical analysis suggests that in contrast to market free of governmental interventions, the PoA in our system decreases.
\end{abstract}


\title{Price of Anarchy in a Double-Sided Critical Distribution System}
\author{David Sychrovsk\'{y}}
\affiliation{
  \institution{Charles University}
  \city{Prague}
  \country{Czechia}}
\email{sychrovsky@kam.mff.cuni.cz}

\author{Jakub \v{C}ern\'{y}}
\affiliation{
  \institution{Nanyang Technological University}
  \country{Singapore}}
\email{cerny@disroot.org}

\author{Sylvain Lichau}
\affiliation{
  \institution{University of Bordeaux}
  \city{Bordeaux}
  \country{France}}
\email{sylvain.lichau@etu.u-bordeaux.fr}

\author{Martin Loebl}
\affiliation{
  \institution{Charles University}
  \city{Prague}
  \country{Czechia}}
\email{loebl@kam.mff.cuni.cz}

\usepackage{todonotes}
\usepackage{algorithm}
\usepackage{algpseudocode}
\usepackage{enumitem}
\usepackage{csquotes}
\usepackage{lipsum}

\usepackage{amsmath,amsfonts,amsthm}
\newtheorem{theorem}[]{Theorem}
\newtheorem{definition}[]{Definition}

\makeatletter
\DeclareRobustCommand*\cal{\@fontswitch\relax\mathcal}
\makeatother

\newenvironment{proofsketch}{%
  \renewcommand{\proofname}{Proof Sketch}\proof}{\endproof}

\def\real{\mathbb{R}}
\def\batch{\mathbb{B}}

\usepackage{framed}
\renewenvironment{shaded}{%
  \MakeFramed{\advance\hsize-\width \FrameRestore\FrameRestore}}%
 {\endMakeFramed}
\definecolor{shadecolor}{gray}{0.85}

\begin{document}

\pagestyle{fancy}
\fancyhead{}


\maketitle

\section{Introduction}

Most of the goods available to the general public are meant to increase the quality of life of individuals or count as luxuries, and are traded using standard market mechanisms. Other resources serve a more social purpose -- when allocated well, they increase the well-being of the entire society like public housing, school seats, or healthcare products. Among those, some are desirable to be readily available to everyone, e.g., essential medicines, various equipment, or even vaccines that enable to reach herd immunity in the population only when enough people have developed protective antibodies against future infections. In times of need like disasters, local epidemics, or even conflicts and wars, these resources need to be distributed swiftly and in a highly organized manner to reach as many eligible people as possible in a limited timeframe.

Allocating such public resources is commonly reserved for governmental services and done at prices below market-clearing or even free of charge. However, leaving the competitive markets out of the allocation process often results in inefficiencies, both economic and temporal, caused by problems inherent to centralized planning~\cite{moroney1997relative}. On the other hand, real-world trading markets, frequently modeled as large double auctions with many sellers and buyers on each side, are capable of distributing the goods flexibly and reliably. The problem remains that even though, with increasing size, the participants are incentivized to be truthful (which leads to asymptotic efficiency~\cite{cripps2006efficiency}), the resulting goods reallocation is not necessarily socially optimal in terms of being available to everyone. Any discrepancy in wealth is then only exacerbated by crises similar to the coronavirus pandemic or the war in Ukraine we experienced in recent years. In such settings, scarce resources necessary for keeping the society up and running could be easily swayed by its more fortunate members, which has to be countered by carefully designed measures.

As an attempt to combine the best of both worlds, the following hybrid distribution system called Crisdis is suggested in ~\cite{crisdis22}: a trustworthy central authority provides a marketplace where buyers and sellers engage in two-sided repeated trading over a period of many days. At the beginning of each market day, the authority allocates buying rights to the participating members (e.g., individual hospitals), which are traded together with the goods. Everything else (pricing, storage, delivery, etc.) is left up to the sellers and buyers themselves, with one requirement only: at the end of each trading day, each buyer needs to possess the number of rights greater or equal to the number of goods. The straightforward motivation for this arrangement is that the traders selling some of their assigned rights obtain extra funds, which they can use in future markets to satisfy their demand for the critical goods better. Another motivation is that the needs of individual participants used to allocate the rights can be evaluated independently by the central authority using real-time crisis data, thus sidestepping the bottleneck of many auction mechanisms -- the proneness to strategic manipulation. 

The utmost priority of Crisdis is to improve the accessibility of critical goods to all eligible buyers during crises in a trading system which is as realistic as possible. For this purpose, \cite{crisdis22} introduced a measure of the social efficiency of the allocations realized by the semi-distributed system called \textit{frustration}. Frustration can be seen as a scaled negative difference between fairness and reality: for a participating buyer, it is the scaled difference between the (potential) allocation of rights to the buyer and the number of goods purchased by them if the value is at least zero, and zero otherwise. Assuming the market attains its equilibrium, the sum of frustrations of the traders describes the system's \textit{Price of Anarchy}, i.e., the price the society pays for allocating the goods through the market and not directly as suggested by the fairness mechanism. 

In \cite{crisdis22}, the authors study how frustration evolves during repeated interactions in the system under a single-sided auction mechanism based on activities of buyers.
This work contributes by a thorough study of a more realistic double-sided mechanism.

\subsection{Contributions}
\label{sub.goal}
We study trading in a system consisting of a sequence of complex double-sided markets combined with a fairness mechanism for allocating the rights designed to improve social good. Following \cite{crisdis22}, we focus on the well-known and thoroughly studied \textit{contested garment distribution}~\cite{AM} for fairly allocating the rights\footnote{Our experimental results show that this fairness mechanism performs well in practice.}, but expand on their work by analyzing multiple double-auction market mechanisms instead of a simple English auction. We introduce these double-auction mechanisms in Section 3, ranging from random acceptable allocations to maximum clearing under average-price bids. 

Our priority in this work is to study the behavior of a large complex system, which makes it difficult to analyze the traders' behavior theoretically. For this reason, in Section 4, we introduce a reinforced-learning algorithm in an attempt to approximate the system's equilibrium. In Section 5 we present the empirical results. First, we perform a thorough numerical analysis demonstrating how close to the equilibrium we are able to converge to. Then we carry out a series of ablation experiments, showing that the Price of Anarchy in the system without the fairness mechanism may be high.
We confirm that together with intuitive governmental regulations akin to increased storing prices for the goods, the system with the fairness mechanism is able to decrease the Price of Anarchy. In the last part of the paper, we summarize the desired features of the trading system enlightened by the experiments.

\subsection{Related work}

Our work belongs to the literature on redistributive mechanisms, especially those mitigating inequalities. Perhaps the most related paper studies a two-sided market trading goods of homogeneous quality, optimizing the traders' total utilities~\cite{dworczak2021redistribution}. The difference lies in our explicit incorporation of the buyers' varying needs into the consideration and the fact that in our model, the utilities are a common knowledge. This work was recently generalized into a setting with heterogeneous quality of tradable objects, more diverse measures of allocation optimality, and imperfect observations about the traders~\cite{akbarpour2020redistributive}. Another related work presents multiple markets and non-market mechanisms for allocating a limited number of identical goods to several buyers~\cite{condorelli2013market}. The author shows that when the buyers' willingness to pay coincides with the designer's allocation preferences, market mechanisms are optimal, and vice versa. In crises environments studied in our work, it is reasonable to assume that the critical resources are highly valuable to all participants, yet, some may lack the money to obtain them. Together with the fact that it is in society's best interests to allocate the goods fairly, these results suggest that leaving the distribution solely to unregulated markets is rather inadvisable.

\section{Problem Definition}

We assume the existence of a centralized marketplace where critical goods are traded periodically during the entire distribution crisis among the buyers and the sellers using an internal currency.  We consider only one type of good and call it Good. To simplify the presentation we assume here that the Good is {\em divisible}\footnote{The results easily generalize to periodic trading of larger quantities of indivisible Good; see \cite{crisdis22} for a more detailed explanation.}. We refer to each trading period of the distribution crisis as a \textit{Market}. A finite sequence of Markets then forms a \textit{Crisis}. The structure of the entire system is depicted in Figure~\ref{fig:crisdis}, and a simple example is available in Appendix~\ref{app: toy example}. 

In order to reduce volatility and promote fairness during trading, similarly as in~\cite{crisdis22}, we introduce a new type of tradable resource called Right. In each Market, in order to buy the Good, the buyer also needs to possess an equivalent amount of the Right. The Rights are allocated to the buyers before the trading in a Market begins by a centralized \textit{Fairness mechanism} using the sellers' declared offers and buyer's declared demands. The traders then engage in a series of interactions resulting in their announcement of bids. A dedicated \textit{Market mechanism} then allocates the Goods and Rights based on the bids. 

The residual resources of Good after consuming the demands, and the Money are then transferred to the next Market, as we model the shortage of the critical Good over an extended period of time. However, the residual of Right disappears after each Market. 

\begin{figure}[t]
    \centering
    \includegraphics[width=.95\linewidth]{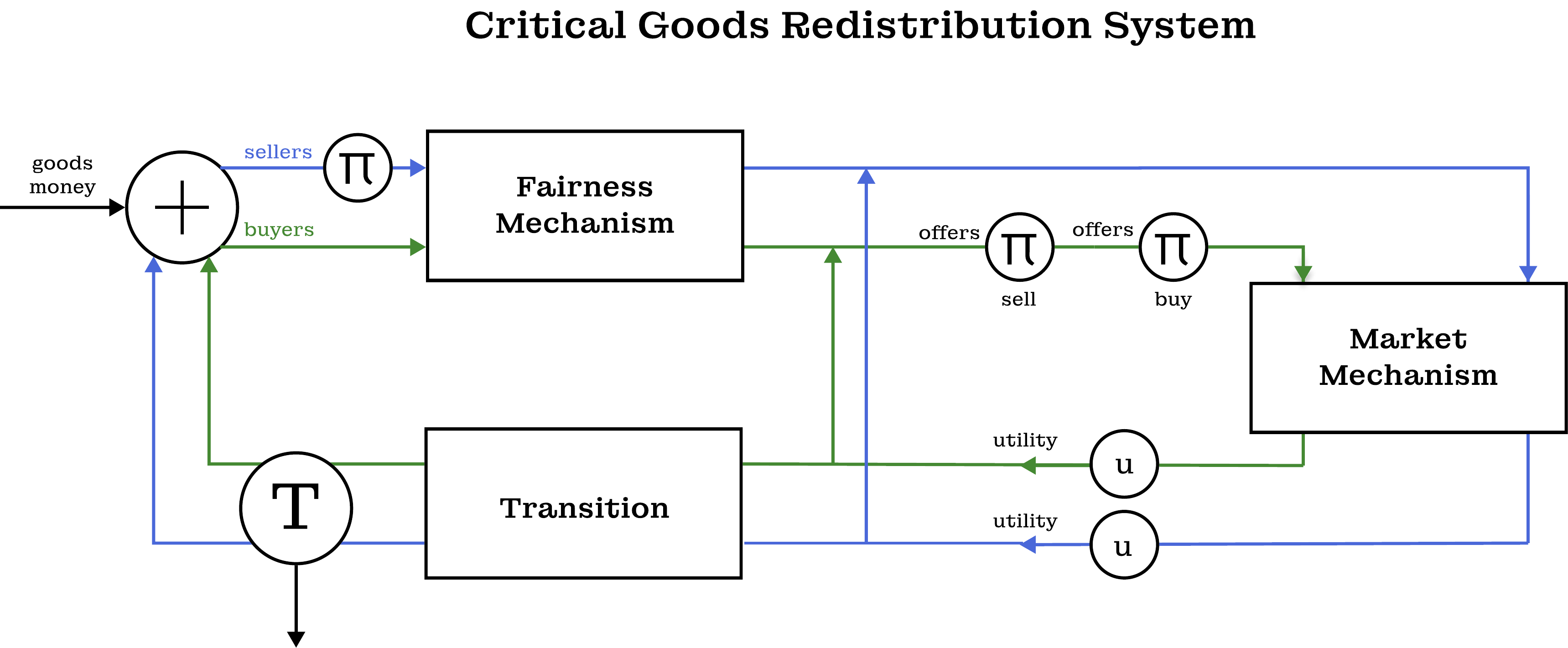}
    \caption{Fairness and market mechanisms positioned in a feedback loop of our redistribution system. One iteration of the outer loop corresponds to one Market. $\pi$ refers to the strategies and $T$ checks the termination condition.}
    \label{fig:crisdis}
\end{figure}

\subsection{One trading period: the Market}\label{subsec: market mechanism}

Formally, we model the trading of the goods during the Market as a double-auction represented as a parametric\footnote{Why this market game and its strategies are parametric in $({\cal M},{\cal G})$ will become apparent later during the definition of the sequence of markets constituting the crisis.} imperfect-information sequential game ${\mathbb{G}}({\cal{M}}, {\cal{G}}) = ({\cal{T}}, {\cal{M}}, {\cal{G}}, {\cal{D}}, \phi, \mu, u, k)$. We use calligraphic letter to denote ordered sets and non-calligraphic letters to denote individual elements of these sets. The set of traders $\cal{T}$ consists of buyers $\mathcal{B}$ and sellers $\mathcal{S}$; the sets of buyers and sellers are assumed to be disjoint.
The set ${\cal{M}}=(M_1,M_2,\dots,M_{|{\cal T}|})\in \mathbb{R}^{+,|{\cal{T}}|}_0$  determines the real and non-negative amount of money each trader receives at the beginning of the Market. Similarly, the set ${\cal{G}}=(G_1,G_2,\dots,G_{|{\cal T}|})\in \mathbb{R}^{+,|{\cal T}|}_0$ specifies the real and non-negative amount of Good each trader obtains.

The demands ${\cal{D}}=(D_1,D_2,\dots,D_{|{\cal B}|})\in \mathbb{R}^{+,|{\cal B}|}_0$ describe the ideal real and non-negative amount of Good each buyer hopes to acquire during the trading. 
Function $\phi$ then implements the fairness mechanism, assigning real-valued, non-negative Rights ${\cal{R}} = (R_1,R_2,\dots,R_{|{\cal B}|})\in \mathbb{R}^{+,|{\cal B}|}_0$ to the individual buyers. 

To allocate the Rights, the mechanism needs to know the amount of Good put up for trade. This amount is given by the parametric strategies of the sellers. For each seller $s\in \mathcal{S}$, the set ${\Pi}_s$ contains the seller's all possible parametric strategies, represented as functions $$\pi_s:\real_0^{+,|\mathcal{B}|}\times\real_0^{+,|\mathcal{B}|}\times\real_0^+\rightarrow\real_0^{+,2},$$ and interpreted as $\pi_s({\cal M}_{\mathcal{B}}, {\cal G}_{\mathcal{B}}, G_s) = (v_s^G, p_s^G)$, where $v_s^G\leq G_s$ is the amount of Good offered at price $p_s^G$. Each ${\pi}_s$ is hence a function of the amount of Good and Money each buyer possesses as well as the amount of Good the seller has, but not of the amount of Good available to the other sellers. This corresponds to sellers investing in some market research\footnote{We are primarily interested in the case where buyers are hospitals. In such a scenario, it would not be difficult to obtain an accurate estimate of the funds and supply. The Rights assigned to each buyer are public information.}.
A profile of one strategy per seller is denoted as ${\pi}_\mathcal{S}\in{\Pi}_\mathcal{S}$. The traders' behavior in a specific game ${\mathbb{G}}({\cal{M}}, {\cal{G}})$ is then written as $\pi_\mathcal{S}({\cal{M}}, {\cal{G}})$. We also refer to their offers at once as $v_\mathcal{S}^G$ and $p_\mathcal{S}^G$. The fairness mechanism is defined as: 

\begin{definition}\label{def:fairness}
For any sellers' strategy profile ${\pi}_\mathcal{S}$, the fairness mechanism is a function $\phi:\mathbb{R}^+_0\times \mathbb{R}^{+,|\mathcal{B}|}_0\rightarrow \mathbb{R}^{+,|\mathcal{B}|}_0$ allocating Rights to each buyers, satisfying in each game ${\mathbb{G}}({\cal{M}}, {\cal{G}})$
\begin{alignat*}{2}
    &\sum_{b\in \mathcal{B}}\phi_b(V, {\cal{D}}) = V
    &&\forall\ {\cal{D}}\in \mathbb{R}^{+,|\mathcal{B}|}_0,\\
    &D_b=0\Rightarrow\phi_b(V, {\cal{D}}) = 0
    &&\forall\ {\cal{D}}\in\mathbb{R}^{+,|\mathcal{B}|}_0,\forall b\in \mathcal{B}\\
    &\phi_b(V, \alpha({\cal{D}})) = 
    \phi_{\alpha^{-1}(b)}(V, {\cal{D}})\hspace{3ex}
    &&\forall\ {\cal{D}}\in\mathbb{R}^{+,|\mathcal{B}|}_0,\forall b\in \mathcal{B},\forall \alpha,
\end{alignat*}
where $\alpha$ is a permutation of $\mathcal{B}$ and $V(\pi_\mathcal{S},{\cal M}, {\cal G}) = \sum_{v_s^G\in{\pi}_\mathcal{S}({\cal{M}}, {\cal{G}})}v_s^G$.
\end{definition}

\noindent In this work, we focus on a fairness mechanism implementing the {\it contested garment distribution} (CGD) \cite{AM} as a well-studied baseline fairness method, and refer to the assigned Rights further on in the text as ${\cal R} = \phi(V(\pi_s,{\cal M}, {\cal G}), {\cal D})$. 

Function $\mu$ is of more importance to us, allocating the resources after the bidding phase. The bidding is determined by the strategies of the buyers, who act in two simultaneous-move phases. First, after observing the seller's offers as well as the amount of Good, Right and Money they have, each buyer declares the amount and price of Right $(v_b^R, p_b^R)$ they are offering for sale. Second, after observing the offers of other buyers, each buyer declares the amount and price of Good and Right $(\overline{v}_b^R, \overline{p}_b^R, \overline{v}_b^G, \overline{p}_b^G)$ they are willing to buy. We assume the announced prices can be interpreted as either maximum or average (if buying from multiple sellers) prices, and we construct different $\mu$ mechanisms for each. A strategy of each seller can be then formulated in both stages at once as a function $$\pi_b:\real_0^{+,|\mathcal{S}|}\times\real_0^{+,|\mathcal{S}|}\times\real_0^+\times\real_0^+\times\real_0^+\times\real_0^{+,|\mathcal{B}|}\times\real_0^{+,|\mathcal{B}|}\rightarrow\real_0^{+,6},$$written as $\pi_b(v_\mathcal{S}^G, p_\mathcal{S}^G,M_b,G_b,R_b,v_\mathcal{B}^R, p_\mathcal{B}^R) = (v_b^R, p_b^R, \overline{v}_b^R, \overline{p}_b^R, \overline{v}_b^G, \overline{p}_b^G)$. 
For each buyer $b\in \mathcal{B}$, their set of strategies is ${\Pi}_b$. A profile of one strategy per buyer is denoted as ${\pi}_\mathcal{B}\in{\Pi}_\mathcal{B}$.
We write the Cartesian product $\Pi_\mathcal{S}\times\Pi_\mathcal{B}$ as $\Pi$ and denote $\pi\in\Pi$ a profile of all traders' strategies. The market mechanism is then formally defined as:

\begin{definition}\label{def:market_mechanism}
For any traders' strategy profile ${\pi}$ in a game ${\mathbb{G}}({\cal{M}}, {\cal{G}})$, the market mechanism is a function $\mu:{\Pi}\times\real_0^{+,|{\cal T}|}\times \mathbb{R}^{+,|\mathcal{B}|}_0\times\real_0^{+,|{\cal T}|} \to \mathbb{R}^{+,|{\cal{T}}|}_0\times \mathbb{R}^{+,|\mathcal{B}|}_0\times \mathbb{R}^{+,|{\cal{T}}|}_0$ which returns a realization of trades, i.e., a reallocation of Good, Right and Money among the traders at the end of the Market. 
We further require that $\mu$ satisfies that 
\begin{enumerate}
    \item no trader sells more Good or Right than they offer;
    \item no buyer buys more Good or Right than they declare;
    \item no trader sells Good or Right for a lower than the asking price;
    \item no buyer buys Good or Right for a higher (or higher on average) price than is their bidding price; and
    \item no buyer buys Rights from themselves.
\end{enumerate}
We abuse the notation a little and write $\mu^G(\cdot)$ and $\mu^M(\cdot)$ to refer to the restrictions to reallocated Goods and Money, respectively.
\end{definition}

\noindent Note that the last condition ensures that the desired amount of Right is actually what a buyer would expect. Without it, the buyers can trade virtually with themselves and thus get a lower amount of Right from the Market, even if they could buy more. The choice of the market mechanism affects the strategizing of the traders to a great extent. We hence dedicate the entire next section to the study of multiple such mechanisms.

What remains is to define the utility functions $u=(u)_{t\in{\cal T}}$, $u_t:\Pi\times\real_0^{+,|{\cal T}|}\times\real_0^{+,|{\cal T}|}\to\real$ for each trader $t$. The sellers are motivated solely by profit. Thus, the utility they get from the Market is the amount of Money they receive. We refine this simple model by {\em adding negative utility} for the Good the seller has at the end of each Market. This penalty represents the societal desire for the sellers to sell most of the available critical Good, and may be implement, e.g., through the state penalties which are usually in place during crises\footnote{Without such penalty and if the Good is not perishable and the distribution crisis continues for a longer time, the strategic behavior of sellers would probably be to keep selling small amounts of the Good for very high prices.}. Moreover, in case the Market terminates the Crisis, the sellers obtain also a small additional utility compensating for the Good they still keep stocked. 
Formally, for each $s\in{\cal S},$
\begin{equation} \label{def.u} 
     u_s(\pi,{\cal{M}}, {\cal{G}}) = 
    \begin{cases}
    \mu^M_s(\pi,{\cal{M}},{\cal R}, {\cal{G}}) + C_1 \mu^G_s(\pi,{\cal{M}},{\cal R}, {\cal{G}}) & \text{NT,}\\
    u_s\text{-NT} +  C_2 \mu^G_s(\pi,{\cal{M}},{\cal R}, {\cal{G}}) & \text{T,}
    \end{cases}
\end{equation}

\noindent where NT/T denote non-terminal/terminal markets, and $C_1$ and $C_2$ are suitable constants. The utility of a buyer should incentivize them to keep a steady supply of Good throughout the Crisis. Therefore, after each trading period, they receive utility for the Good they have (up to their demand), which represents their regular consumption (e.g., per day). The buyers also receive some small utility $C_3$ per unit of Money they have at the end the Crisis. Formally, for $b\in{\cal B},$
\begin{equation} 
    u_b(\pi,{\cal{M}}, {\cal{G}}) = 
    \begin{cases}
    \min\left\{D_b, \mu^G({\pi},{\cal{M}},{\cal R}, {\cal{G}})\right\} & \text{NT,} \\
    u_b\text{-NT} + C_3 \mu^M({\pi},{\cal{M}},{\cal R}, {\cal{G}}) & \text{T.}
    \end{cases}
\end{equation}
\noindent  The semantic meaning of the additional constants in terminal-state utilities is to compensate for the possible continuation of the Crisis.

Now let us reiterate and describe again the entire process of how sellers and buyers engage in trading in our two-sided market:
\begin{shaded}
\begin{enumerate}[leftmargin=.3cm]
    \item[]\hspace{-.45cm}\textbf{Sellers' stage:}
    \item Each seller declares the amount and price of Good for sale. 
    \item Each buyer is assigned Rights by the {fairness mechanism} $\phi$.
    
    \vspace{.2cm}
    \item[]\hspace{-.45cm}\textbf{Buyers' stage:}
        \item Each buyer declares the amount of Right they are willing to sell along with the asking price.
        \item Each buyer, given the available amounts and asking prices of Good and Right, declares their {\it bidding price and desired amount} of Good and Right, separately.
    \item The bids are cleared using the {market mechanism} $\mu$.
    \item The traders receive their utilities. 
\end{enumerate}
\end{shaded}

\noindent Note that an important aspect of our model is that the buyers can use the Money they obtained only in the next Market of the sequence. This gives the active buyers advantage of buying the critical Good earlier than the passive buyers; the price of this advantage is the cost of buying additional rights.

\subsection{Sequence of Markets: the Crisis}

We assume that trading takes place periodically, in a finite sequence of Markets, denoting, e.g., trading days. Formally, the Crisis is a partially observable stochastic game with continuous state space and continuous action space, denoted as $\mathbb{C} = ({\cal{T}}, {\cal{D}}, \phi, \mu, u, {\cal M}^1, {\cal G}^1, T, P)$. ${\cal{T}}, {\cal{D}}, \phi, \mu$ and $u$ have the same meaning as in our definition of the Market game and are assumed to be fixed throughout the entire Crisis. Each state in this stochastic game corresponds to a Market game, with the initial game being $\mathbb{G}({\cal M}^1, {\cal G}^1)$. We model a full-blown crisis\footnote{We leave the study of boundary situations for future work.}, and assume all traders are stationary-Markovian, basing their strategies on their observations in the Crisis' current state, i.e., a game $\mathbb{G}$ with some parameters $({\cal M}, {\cal G})$. The parametric strategies introduced in the previous section are hence employed in the entire Crisis. After each non-terminal trading, the sellers keep the unsold amount of Good and the buyers keep the Money and unconsumed amount of Good for the next Market. In contrast, the unused amount of Right is disposed of after each Market terminates. The traders also receive additional Goods or Money before the next Market starts. This process is described via the transition function $P:\Pi\times\real_0^{+,|{\cal T}|}\times\real_0^{+,|{\cal T}|}\to\real_0^{+,|{\cal T}|}\times\real_0^{+,|{\cal T}|}$. Using the same notation as for the market mechanism $\mu$, we write
\begin{align*}
    P^G_t(\pi, {\cal M}, {\cal G}) &= \mu_t^G(\pi,{\cal M},{\cal R}, {\cal G}) +
    \begin{cases}
    {G}^1_{t} &\text{if~} t\in \mathcal{S}\\
    0 &\text{if~}t\in \mathcal{B}\text{, and}
    \end{cases}\\
    P^M_t(\pi, {\cal M}, {\cal G}) &= 
    \begin{cases}
    0 &\text{if~} t\in \mathcal{S}\\
    \mu_t^M(\pi,{\cal M},{\cal R}, {\cal G}) +{ M}^1_{t} &\text{if~} t\in \mathcal{B}.
    \end{cases}
\end{align*}
The finite horizon is denoted $T$. As is usual, a situation in which no trader has an incentive to unilaterally change their strategies is called an {\em equilibrium}. We consider optimality under discounting $\gamma$.

\begin{definition}\label{def: equilibrium}
Let $\pi\in\Pi$ induce a series of Markets $(({\cal M}^1, {\cal G}^1), \dots,$ $({\cal M}^T, {\cal G}^T))$. We call $\pi$ an equilibrium if for any unilaterally deviating profile $\hat{\pi}\in\Pi$ and the corresponding series of Markets $((\hat{\cal M}^1, \hat{\cal G}^1), \dots,$ $(\hat{\cal M}^T, \hat{\cal G}^T))$ it holds for all traders $t$ that
\begin{equation*}
    \sum_{i=1}^T \gamma^i u_t(\pi, {\cal M}^i, {\cal G}^i)\geq \sum_{i=1}^T \gamma^i u_t(\hat{\pi}, \hat{\cal M}^i, \hat{\cal G}^i).
\end{equation*}
\end{definition}

\subsection{Frustration and the Price of Anarchy}

Traditionally, the efficiency of a designed system is measured via the notion of Price of Anarchy, which characterizes how much an equilibrial state reached by self-interested decision-makers differs from an optimal state.  The optimality criterion the system's planner usually considers is the Pareto-optimal social-welfare, i.e., an optimal state maximizes the sum of individual players' utilities.

In our work, we borrow this idea, but consider a slightly different objective of the planner. In crises, instead of blindly maximizing all utilities, we are more interested in maintaining an egalitarian society. We hence study how the amount of Good acquired by buyers evolves for different Market mechanisms and compare it to the amount of Rights assigned to them. The resulting discrepancy describes the inherent inequality in the system, formally defined as \textit{frustration}, and serves to measure (conceptually) the same quantity as in traditional Price of Anarchy: the price the system pays for freedom of agents to choose, instead of planning centrally. 

\begin{definition}\label{def:frustration}
The frustration of buyer $b$ in Market $({\cal M}, {\cal G})$ under profile $\pi\in\Pi$ and allocated rights ${\cal R} = \phi(V(\pi_\mathcal{S}, {\cal M}, {\cal G}),{\cal D})$ is then 
\begin{equation*}
    f_b(\pi, {\cal M}, {\cal G}) = \max\left\{\frac{{ R}_b-\mu_b^G(\pi, {\cal M}, {\cal R}, {\cal G})}{{ R}_b}, 0\right\}.
\end{equation*}
\end{definition}

Our Price of Anarchy in the system is then the normalized accumulated frustration the buyers experience in the sequence of $\tau\leq T$ Markets when the equilibrium $\pi$ is reached and induces a series of Markets $(({\cal M}^1, {\cal G}^1), \dots,$ $({\cal M}^\tau, {\cal G}^\tau))$, i.e.,
\begin{equation}\tag{PoA}
    PoA^\tau = \frac{\sum_{i=1}^\tau\sum_{b\in \mathcal{B}} f_b(\pi, {\cal M}^i, {\cal G}^i)}{\tau|\mathcal{B}|}.
\end{equation}

\section{Market Mechanisms}
\label{sub.clear}
In this section, we study how to reallocate the resources in the Market by introducing four mechanisms that schedule individual trades based on the inputs (bids) of the traders. 
Formally, clearing constraints (compatibility of asking and bidding prices and possibly other constraints) will be represented by two bipartite graphs: $G_G= (\mathcal{B},\mathcal{S},E_G)$ which represents the compatibility for trading the Good and 
$G_R= (\mathcal{B}_S,\mathcal{B}_B,E_R)$ which represents the compatibility for trading the Right. Here, $\mathcal{B}_S$ and $\mathcal{B}_R$ are disjoint copies of $\mathcal{B}$, $\mathcal{B}_S$ represents the sellers of Right and $\mathcal{B}_R$ represents the buyers of Right. A trader of $\mathcal{B}$ can be both a seller and a buyer of Right, but $G_R$ does not connect their representing vertices by an edge. Both $G_G, G_R$ are equipped with a positive real {\em weight} $w_G:V_G\rightarrow \real$ and  $w_R:V_R\rightarrow \real$. The weights of the vertices naturally represent the individual amount (of Good or Right) offered for sale and the individual amount (of Good or Right) desired to buy.

We primarily focus on {\em absolute} mechanisms, i.e. those that prohibit {\it any} trades where the bidding price is larger than the asking price, i.e., $\overline{p}_b^G>p_s^G$ and/or $\overline{p}_b^R > p_{b'}^R$. At the end of this section, we introduce a mechanism which relaxes this condition to hold for {\em average} prices.

\subsection{Random allocation}

A simple random trading mechanism used for purchasing both Rights and Goods proceeds as follow. First, the buyers are randomly permuted. In this order, each buyer is given randomly permuted lists of offers of the traders for Good and Right, respectively. A buyer first trades Good with sellers in order given by the list, until he has no Right left. In the second stage, the buyer trades Good and Right in equal amount, again following the list. This continues until they buy in total their desirable volume, or there are no more offers. We also ensure at every step that the asking price is lower than their acceptable price, and the buyer purchases amount up to the amount offered by the other party.

This mechanism has a unsatisfactory property. Since the buyers are presented with offers in random order, they often do not buy the cheapest option. This can be realistic since no single buyer will be able to see all the offers and choose among them. However, if the trading proceeds sequentially, it is natural for the buyer to consider the cheapest offers first. This also gives incentive to the sellers and traders to make offers at a lower price. 

\subsection{Greedy allocation}

This algorithm is a modification of the random allocation which aims to address the issues mentioned in the previous paragraph. At the beginning, the buyers are sorted by the acceptable price of Good $\overline{p}_b^G$ in descending order. The mechanism has again two stages for each buyer. In the first stage, a buyer uses the Rights allocated to them to buy Goods, starting with the cheapest offer. When they have no Right left, they buy the same amount of Rights and Goods, again starting with the cheapest offers for both. We proceed until there is no offer left, or the buyer bought their desirable volume and continue with the next buyer.

Note that the random and greedy allocations are heuristics that are easily implementable  but do not necessarily lead to optimal allocation which clears a maximum amount of bids. 

\subsection{Maximum clearing using absolute prices}
An allocation clearing maximum amount of bids where we also require that no Right is bought without buying equal amount of Good, can be obtained using network flows. We call a mechanism utilizing this approach {\em Maximum clearing}. Its advantage is that it works also for indivisible Good. Another advantage is that the result of the Maximum clearing allocation is the list of individual tradings with compatible asking and bidding prices. The final price of each individual trading may be chosen in various ways from this compatibility interval. 

\begin{theorem}
\label{thm.max}
Maximum clearing allocation can be found efficiently using a reduction to the Max Flow problem. As a consequence, a Maximum clearing allocation is polynomial for both divisible and indivisible Good.
\end{theorem}
\begin{proofsketch}
Construct the network of the Max Flow problem combining $G_G$ and $G_R$. 
See Appendix~\ref{app: proofs} for full version.
\end{proofsketch}

\subsection{Maximum clearing using average prices}
In contrast to the Market mechanisms presented so far, in this variant of the maximum clearing mechanism, we view the prices $\overline p^G_b$ and $\overline p^R_b$ as maximum \textit{average} prices $b$ is willing to pay. This is a relaxation of the maximum clearing mechanism with absolute bids, as it allows for trades which were previously infeasible. Again, the solution can be computed efficiently.

\begin{theorem}
\label{thm.max2}
Maximum clearing allocation with average bids can be found efficiently using a linear program.
\end{theorem}
\begin{proofsketch}
The linear program maximizes the volume of Good sold, while ensuring the average selling price of Good and Right is at most the desired price. The linear program can be solved in polynomial time. 
See Appendix~\ref{app: proofs} for the full proof.
\end{proofsketch}

\begin{shaded}
\noindent\textbf{Repeated buyers' stage.} 
To further improve clearing during one iteration of a Market, we present a simple extension where the \textit{buyers' stage} is repeated $k$-times using the same strategy profile $\pi$. In each round the Goods, Rights, and Money are reallocated according to a selected Market mechanism. 
\end{shaded}

\section{Learning the Game's Equilibrium}

\begin{algorithm}[t]
\caption{Equilibrium Learning Algorithm}\label{alg:learning_alg}
\begin{algorithmic}[1]
\State $\mathcal{B} \gets \textit{set of buyers}, \mathcal{S} \gets \textit{set of sellers}$, $\batch\gets \{\}$
\For{$episode \in \{1,\dots N_\text{sims}\}$}
\For{$\tau\in\{1,\dots T\}$}
\State $G_s \gets G_s + G_s^1$, $M_b \gets M_b + M_b^1$
\State $o_\mathcal{S} \gets$ observation of sellers
\State $\pi_\mathcal{S} \gets \text{clip}(\Pi_\mathcal{S}(o_\mathcal{S}(s)),0,1)$ 
\State Allocate Right according to $\phi$
\State $o_\mathcal{B} \gets$ observation of buyers
\State $\overline{\pi}_\mathcal{B} \gets \text{clip}(\Pi_\mathcal{B}(o_\mathcal{B}(b), \pi_\mathcal{S}),0,1)$ 
\State $\pi_\mathcal{B} \gets \text{clip}(\Pi_\mathcal{B}(o_\mathcal{B}(b), \pi_\mathcal{S}, \overline{\pi}_\mathcal{B}),0,1)$
\State Trade according to a market mechanism $\mu$
\State Compute utilities $u_\mathcal{B}, u_\mathcal{S}$
\State $\batch\gets \batch\cup\{o_\mathcal{B}, o_\mathcal{S}, \pi_\mathcal{S}, \pi_\mathcal{B}, u_\mathcal{B}, u_\mathcal{S}\}$
\State $\mathcal{G}_\mathcal{B} \gets \max(\mathcal{G}_\mathcal{B} - \mathcal{D}_\mathcal{B}, 0)$
\If{$\tau\ \text{mod}\ N_\text{train}$ is zero}
    \State Sample $batch\sim\batch$
    \State Train on $batch$ using TD3
\EndIf
\EndFor
\State Reset episode 
\EndFor
\end{algorithmic}
\end{algorithm}

In this section, we describe a reinforced learning algorithm we use to obtain an approximation of the equilibrium of the Crisis.
We treat the entire interaction as a multi-agent reinforcement learning (MARL) problem as it is common in the literature~\cite{fu2022actorcritic,liu2022neupl,perolat2022mastering,muller19}, with the assumption that the learning algorithm shall converge to a solution close to the equilibrium. We further verify the quality of the solution by computing its exploitability~\cite{nashconv}.
The sellers and buyers are represented as agents who interact in the environment described in sections 2 and 3. Each agent is trained to maximize their own expected discounted future utility in this environment.

\subsection{Utilities in the learning environment}
Next, we focus on the traders' utilities. To identify them, we need to specify constants $C_i$. For simplicity and to reduce the action space, we assume there exists a maximum price $\overline{P}$ the Good and Right can be offered at. Since the offered volume is bounded by the volume owned by a trader, the traders' actions fall in a closed interval. 

Let us focus on the sellers first: their utility is given by two constants representing the price of storing the Good, and the expected future utility for the amount of Good in the terminal Market. We set the latter to be the market clearing price. This means the sellers expect to sell the Good for at least that price, which is a reasonable assumption during a crisis. The price of storing, $C_1$, may be chosen arbitrarily; however, it needs to be sufficiently high. If $|C_1|\frac{T}{2} < C_2$, it becomes beneficial for the sellers to keep the Good, and the selling price would thus be $\overline{P}$.

The buyers' utility is given in terms of the future expected utility for Money in the terminal Market. The relative penalty influences the mean utility a buyer obtains and again, it may be chosen arbitrarily. The future utility for Money is the utility for Good attainable with that Money, which is at least the utility for Good purchased at the maximum price $\overline{P}$. $C_3$ should hence inversely depend on $\overline{P}$.

\begin{figure}[t]
    \centering
    \includegraphics[width=0.35\linewidth]{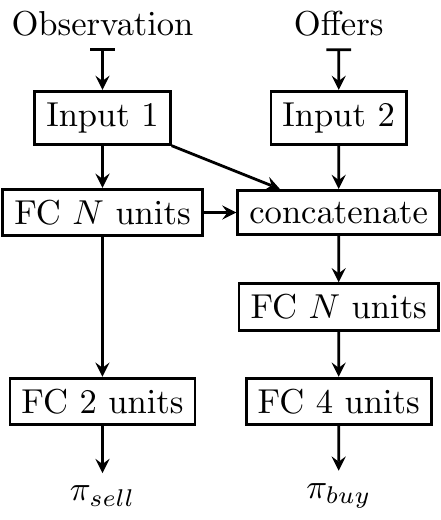}\hspace{0.06\linewidth}%
    \includegraphics[width=0.17\linewidth]{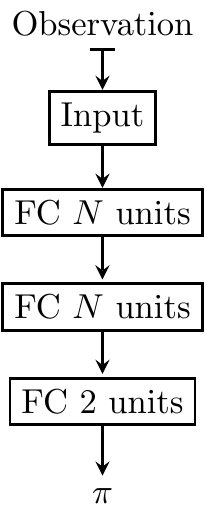}\hspace{0.06\linewidth}%
    \includegraphics[width=0.35\linewidth]{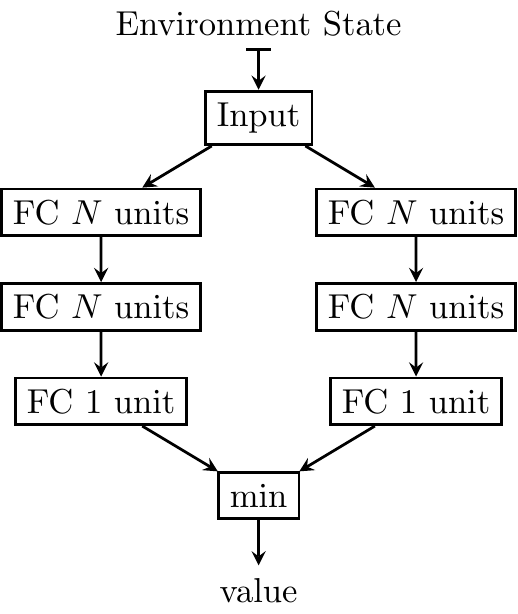}
    \caption{Architectures of the used neural networks: (Left) sellers' actor, (Middle) buyers' actor, and (Right) the critic.}
    \label{fig:neural_architecture}
\end{figure}

\subsection{Learning algorithm and model architecture}
\label{subsec: architecture_and_hyperparameters}

For training the agents' strategies we adopt an {\it actor-critic} algorithm called Twin-Delayed Deep Deterministic Policy Gradient (TD3)~\cite{TD3}. We modify it for the purpose of finding a solution in our scenario and depict the pseudocode in Algorithm~\ref{alg:learning_alg}. The policy $\Pi_t$ of each trader $t$ is a random variable with a Gaussian distribution with mean and standard deviation parameterized by a neural network.

\begin{figure*}[t] %
\centering
\includegraphics[width=.24\textwidth]{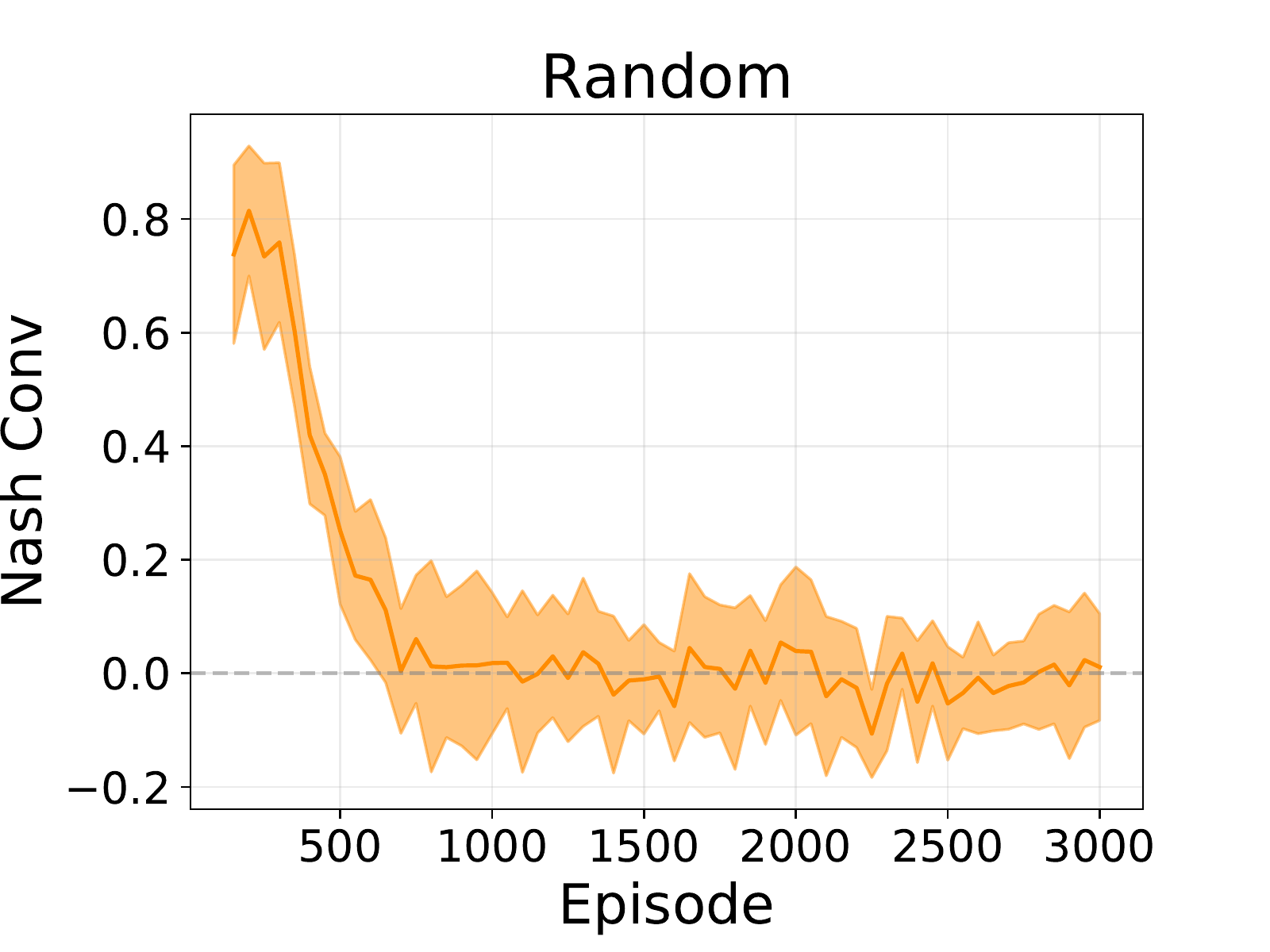} %
\includegraphics[width=.24\textwidth]{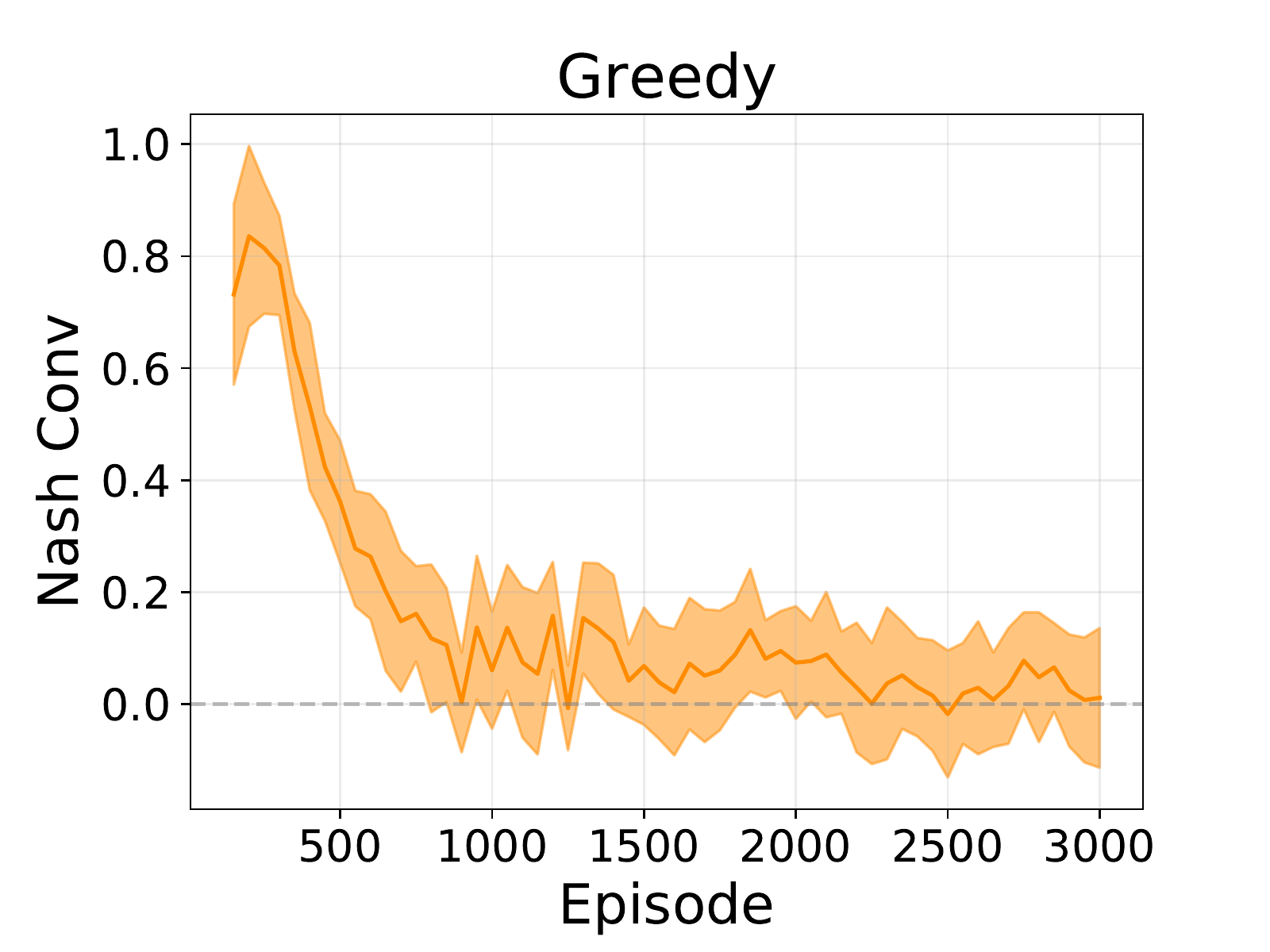} %
\includegraphics[width=.24\textwidth]{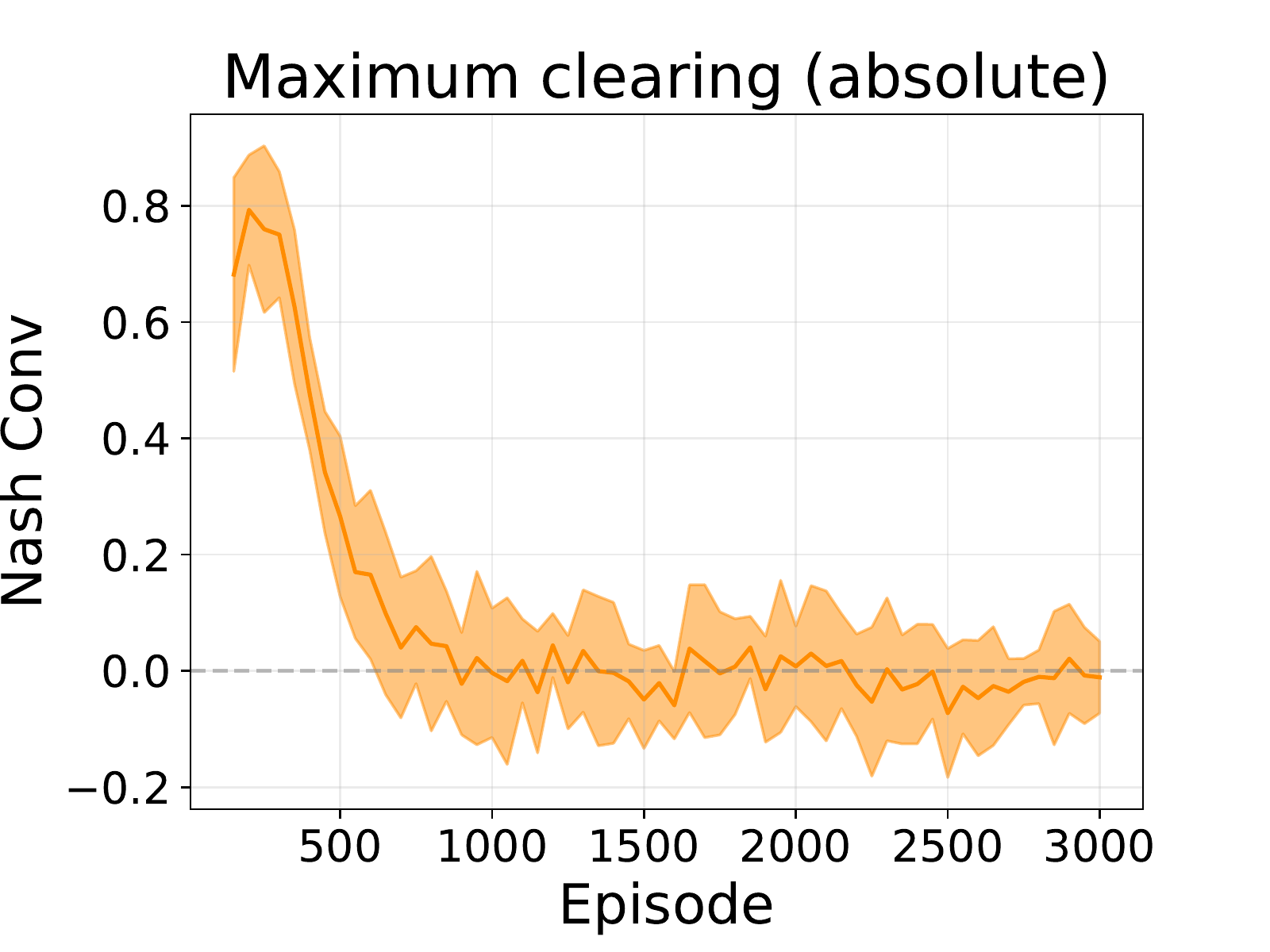} %
\includegraphics[width=.24\textwidth]{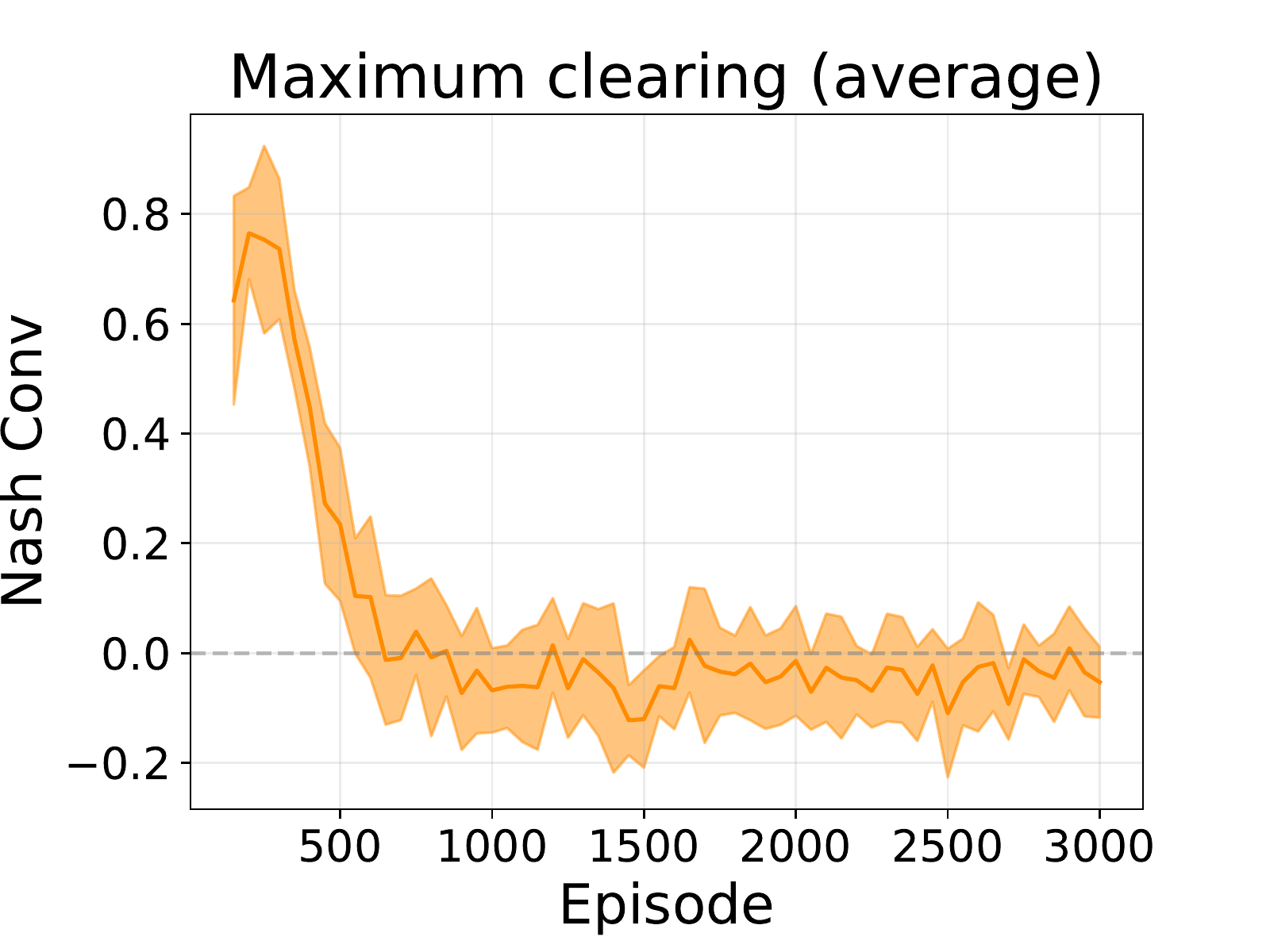} %
\caption{The exploitability of candidate solutions when learning the equilibria in systems with Rights and $k$=1 trading period for four different market mechanisms. The graphs show means and standard deviations over ten random seeds.%
}
\label{fig: nashconv}
\end{figure*}

The architecture of neural networks we employ is shown in Figure~\ref{fig:neural_architecture}. The buyers' actor needs to process the offers of the sellers and consecutively offer the Right for sale before processing the offers of other buyers. To accomplish that, the output of the first hidden layer is concatenated with the offers of the other buyers, and only the first hidden layer is used to predict the buyer's offer. In this way, the network can be used to obtain the buyer's offer without the offers of others influencing the result. Since the actions come from a bounded interval, the actors use a sigmoid activation function on the output layer on the means, which is then properly rescaled. The standard deviation uses the softplus activation.

Moreover, we enhance the vanilla TD3 algorithm with upgoing policy update \cite{alpha_star} and reward clipping to $[-1, 1]$. To accelerate training, we allow the sellers to share the same replay buffer $\batch$. This makes sellers' policies similar without using an identical actor.

\section{Empirical Evaluation}\label{sec:experimental_results}

Finally, we demonstrate the properties of our hybrid system with fairness and market mechanisms, and the effectiveness of our learning algorithm, on practical examples. First, we assess the quality of the learned solutions using NashConv, a measure of exploitability. In the second part, we study how the learning algorithm scales with the number of traders. Lastly, we analyze to which degree the incorporation of Rights affects the Price of Anarchy of the approximated equilibrium throughout the entire crisis. 

We evaluate systems combining three degrees of fairness with all four market mechanisms from Section~\ref{sub.clear}. The variants of fairness we consider are systems with: (i) no distributed Rights (i.e., a free market), (ii) Rights and $k$=1 tradings; and (iii) Rights and $k$=2 tradings.

\paragraph{Experimental setting} All experiments were conducted on a computational cluster with AMD EPYC 7532 CPUs running at 2.40GHz. We utilized only 5 of its 16 cores and 3GB of RAM. The code was implemented in Python using tensorflow 2.6, tensorflow-probability 0.15, mip 1.14, and numpy 1.21. The open-source CBC solver carried out all LP computations. The complete list of all hyperparameters of Algorithm~\ref{alg:learning_alg} can be found in Appendix~\ref{app:hyp}, and the code \href{https://github.com/DavidSych/Price-of-Anarchy-in-a-Double-Sided-Critical-Distribution-System}{here}.

\paragraph{Experimental domain} We consider a sequence of $T=10$ Markets with four buyers and four sellers. We choose a prototypical setting where three of the four buyers receive significantly more funds then the last buyer. At the same time, this last buyer suffers a large demand, in most cases exceeding the demands of the others. We refer to the first three buyers as \textit{rich} and to the last buyer as \textit{poor}. We generate the instances of this setting by sampling the demands and the earnings of the buyers uniformly randomly from given intervals. For the rich buyers, the demand $D_b$ in drawn from $\mathcal{U}(1, 2)$ and the earning $M^1_b\sim\mathcal{U}(4, 6)$. For a poor buyer, $D_b\sim\mathcal{U}(4, 6)$ and $M^1_b\sim\mathcal{U}(1, 2)$. The set of demands and earnings is then normalized such that $\mathbb{E}_{b\sim \mathcal{B}}[D_b] = 1$ and $\mathbb{E}_{b\sim \mathcal{B}}[M^1_b] = {1}/{8}$. To fix a scale\footnote{This corresponds to choosing a currency s.t. the price of a unit of Good is at most 1.}, we set the maximum price as $\overline{P}=1$. The constants in the utilities are then chosen as $C_1=-1 / 8$, $C_2={1}/{2}, C_3 = {1}/{\overline{P}} = 1$ and $G_s^1= 1 / |\mathcal{S}|$.

\begin{figure}[t!] %
\centering
\includegraphics[width=.49\linewidth]{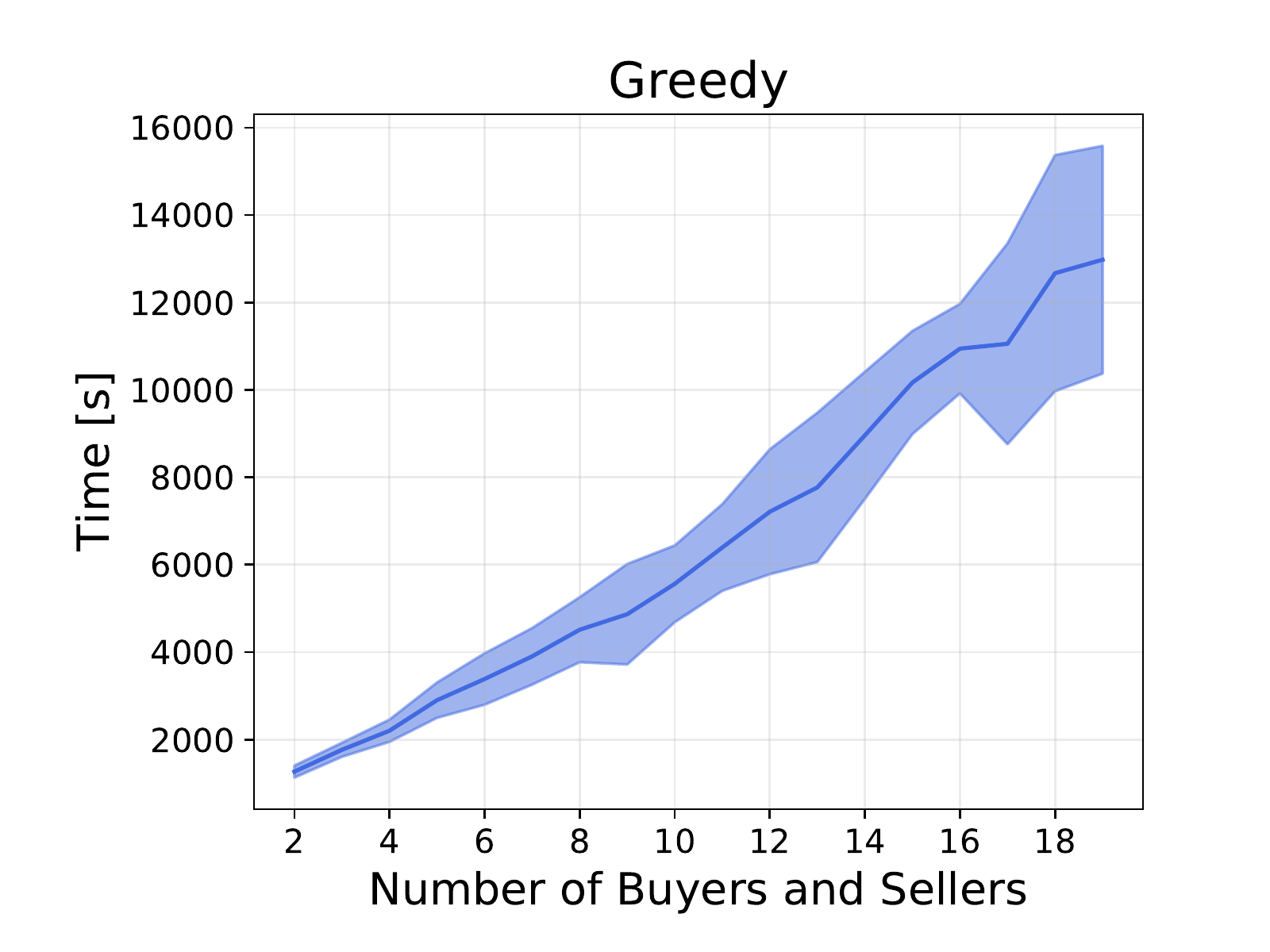} %
\includegraphics[width=.49\linewidth]{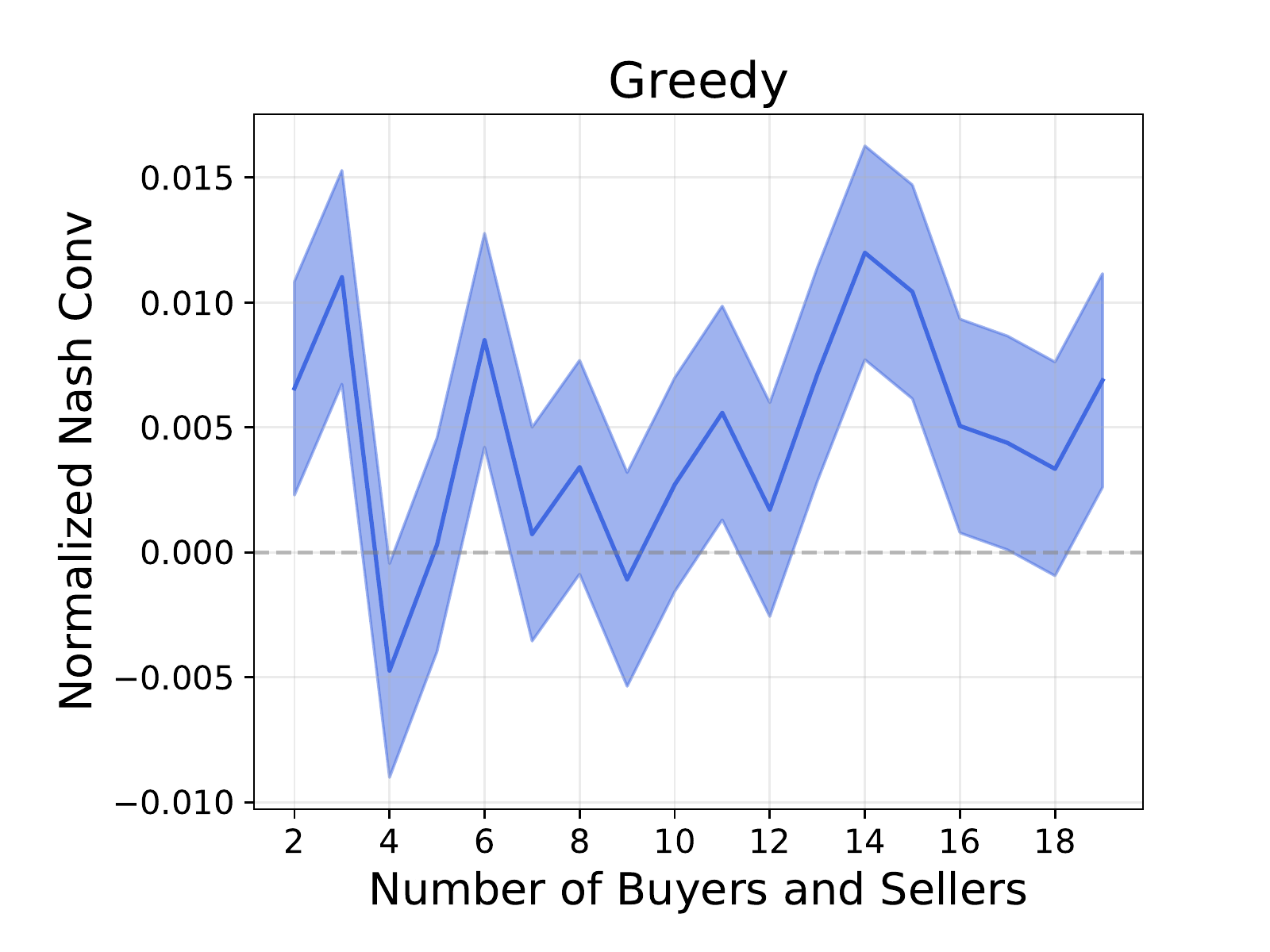} %
\caption{Computational time to learn equilibria and normalized final exploitability as functions of number of traders.}
\label{fig: time complexity}
\end{figure}

\subsection{Exploitability}\label{subsec: convergence}

We measure the quality of a candidate solution from episode $\tau$ through its exploitability. For computing the exploitability we employ the notion of NashConv \cite{nashconv}, given as
\begin{equation*}
    \sum_{i=1}^T\sum_{t\in\mathcal{T}} u_t(\pi^\tau_{-t}\cup\overline{BR}_t(\pi^\tau), \overline{\cal M}^{i_t}, \overline{\cal G}^{i_t}) - u_t(\pi^\tau, {\cal M}^i, {\cal G}^i).
\end{equation*}
Here, we denote by $\overline{BR}_t(\cdot)$ an approximate best response of trader $t$. The complement strategies are then written as $\pi_{-t}$. The corresponding sequences of Markets for $\pi^\tau$ and $\pi^\tau_{-t}\cup\overline{BR}_t(\pi^\tau)$ are denoted as $({\cal M}^i, {\cal G}^i)$ and $(\overline{\cal M}^{i_t}, \overline{\cal G}^{i_t})$, respectively. We train a best-response of each trader separately for 100 episodes, keeping the opponents' policies fixed, and starting from the policy of trader $t$ in $\pi^\tau$. 
In Figure~\ref{fig: nashconv} we present the results achieved with all four market mechanism in a system with Rights and $k=1$, averaged over 10 runs with different random seeds. The results suggest the algorithm was able to reach a sufficiently close approximation of the equilibrium. Moreover, we verified the inclusion of Rights or the value of $k$ do not have a significant effect on exploitability.

\begin{figure*}[t!] 
\centering
\includegraphics[width=.24\textwidth]{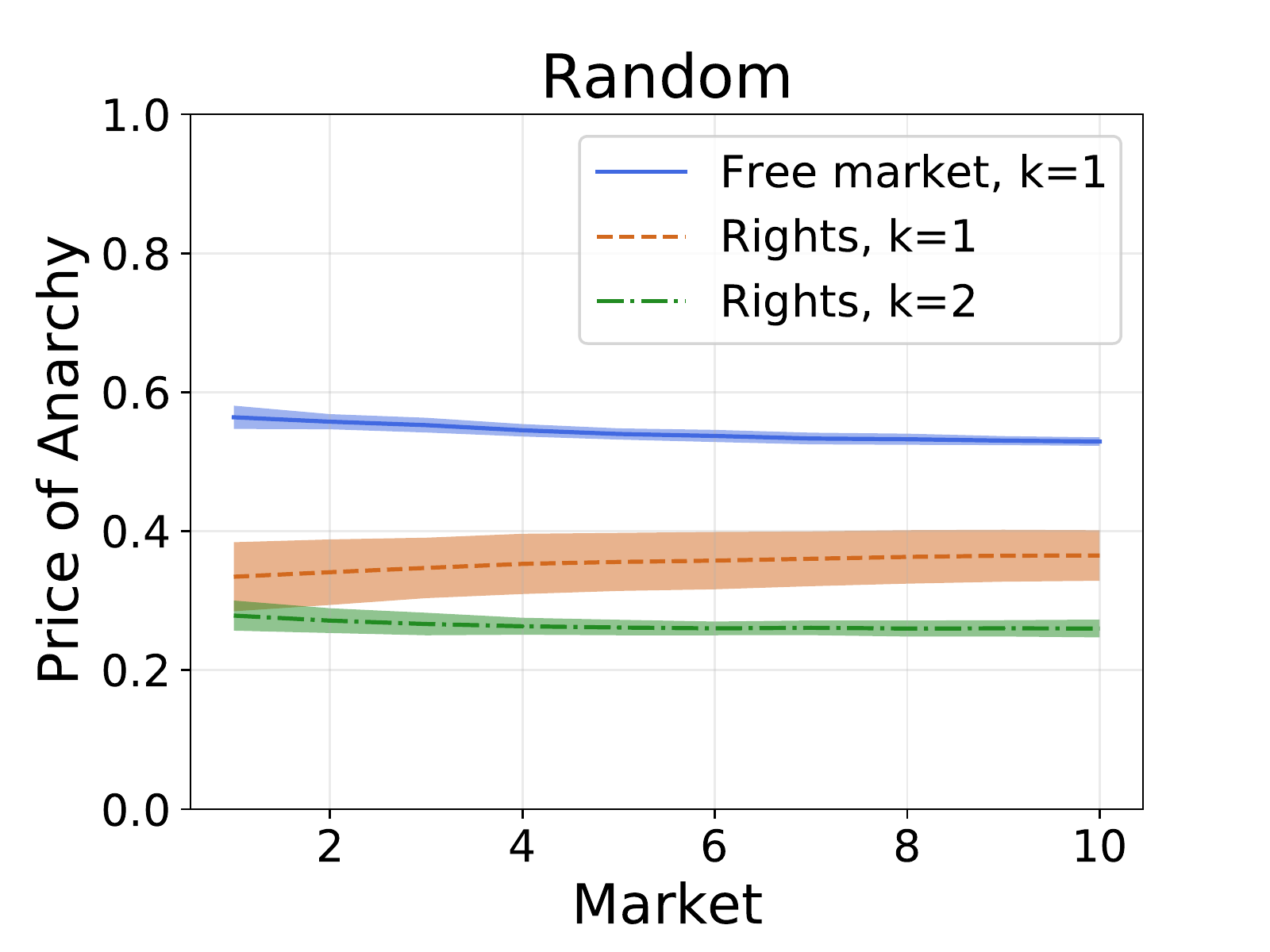} 
\includegraphics[width=.24\textwidth]{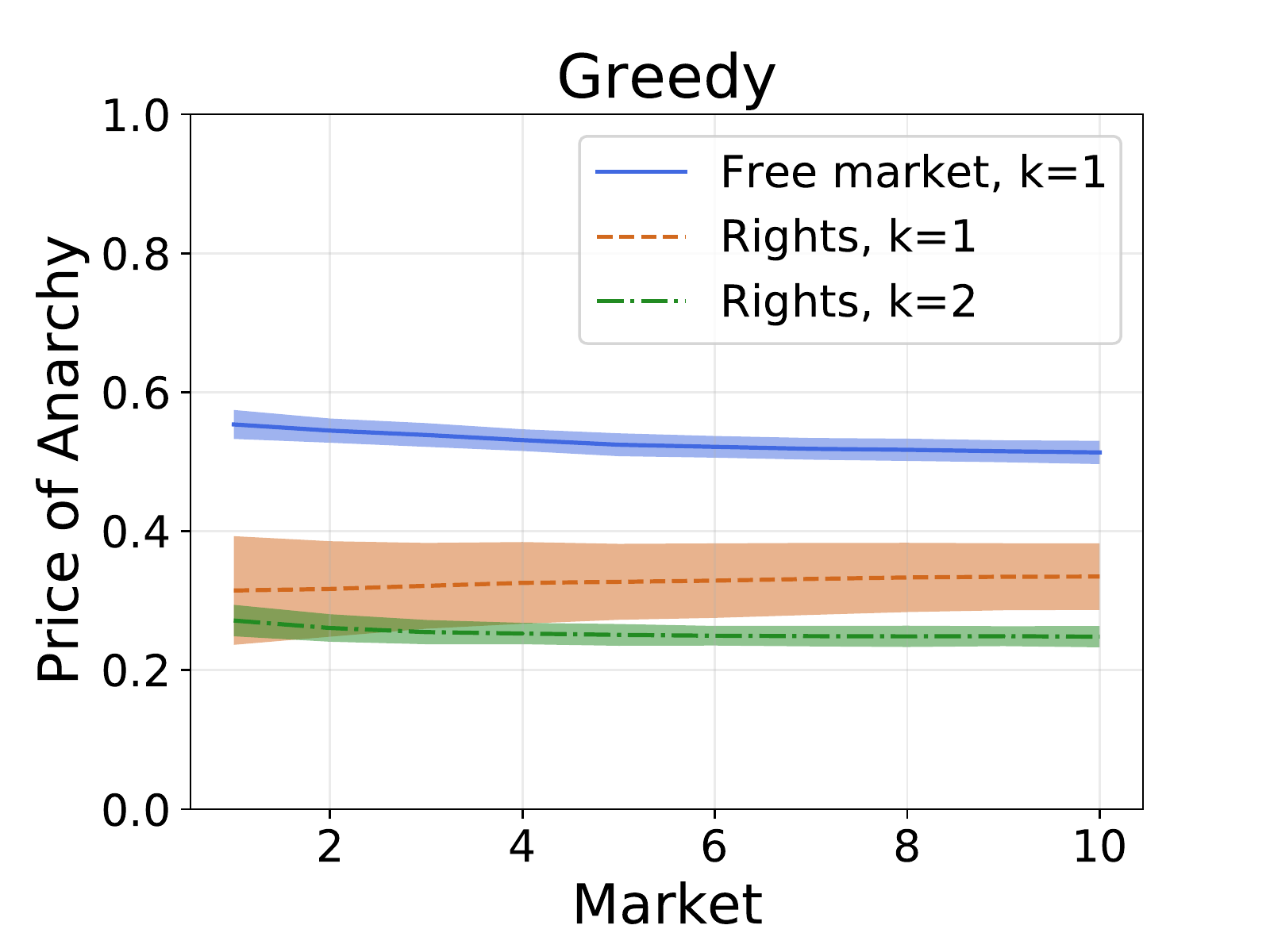} 
\includegraphics[width=.24\textwidth]{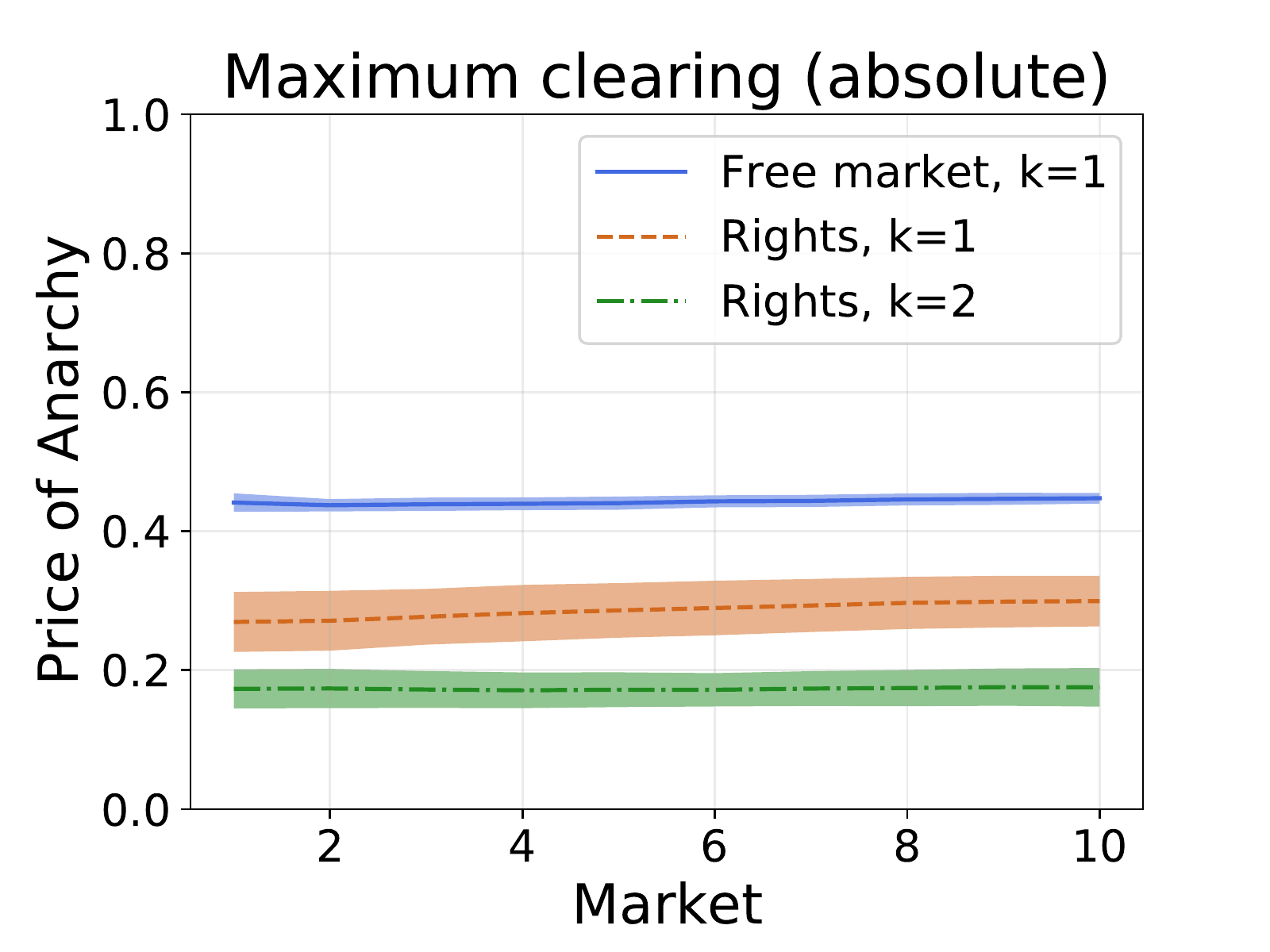} 
\includegraphics[width=.24\textwidth]{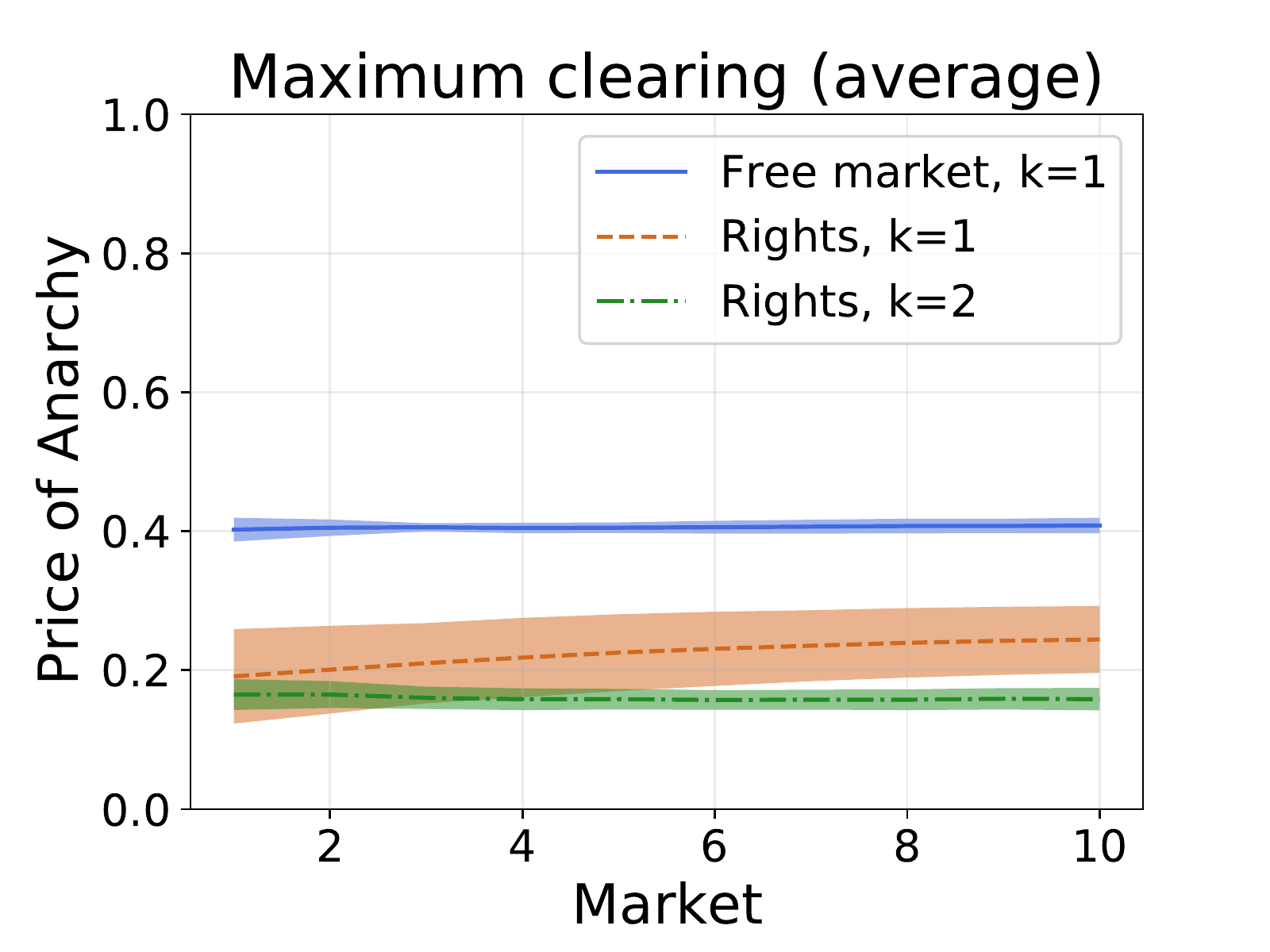} 
\includegraphics[width=.24\textwidth]{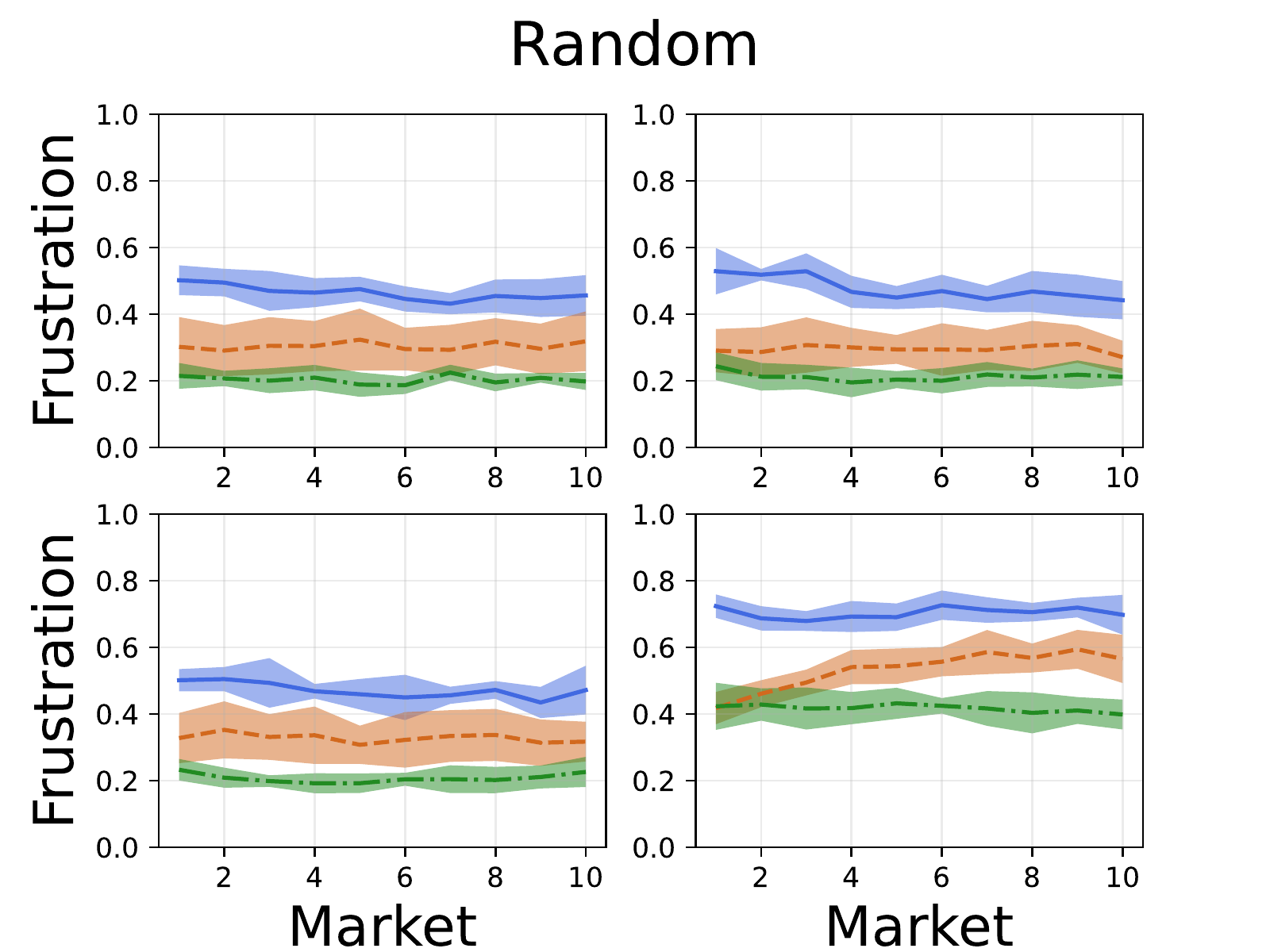} 
\includegraphics[width=.24\textwidth]{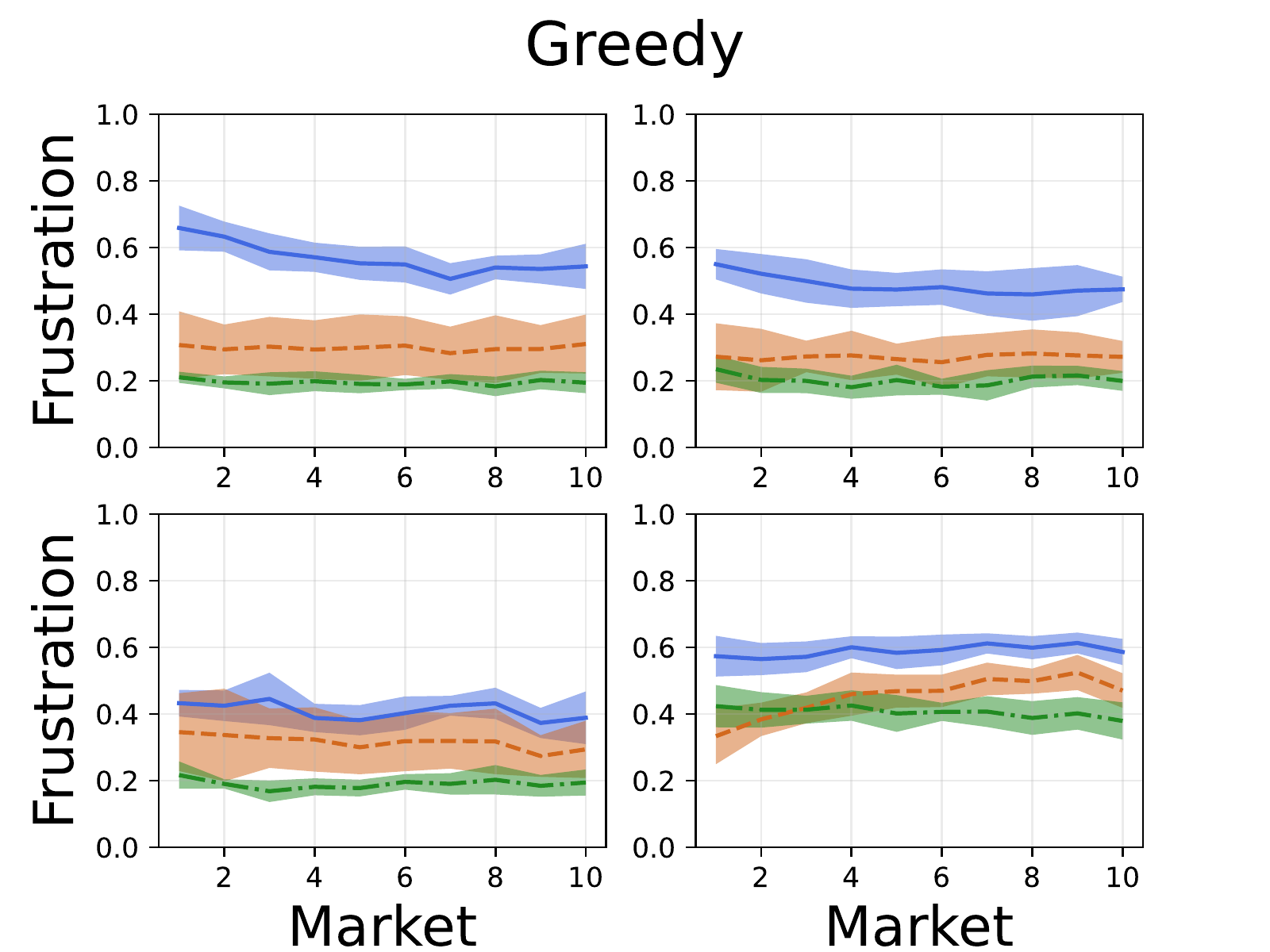} 
\includegraphics[width=.24\textwidth]{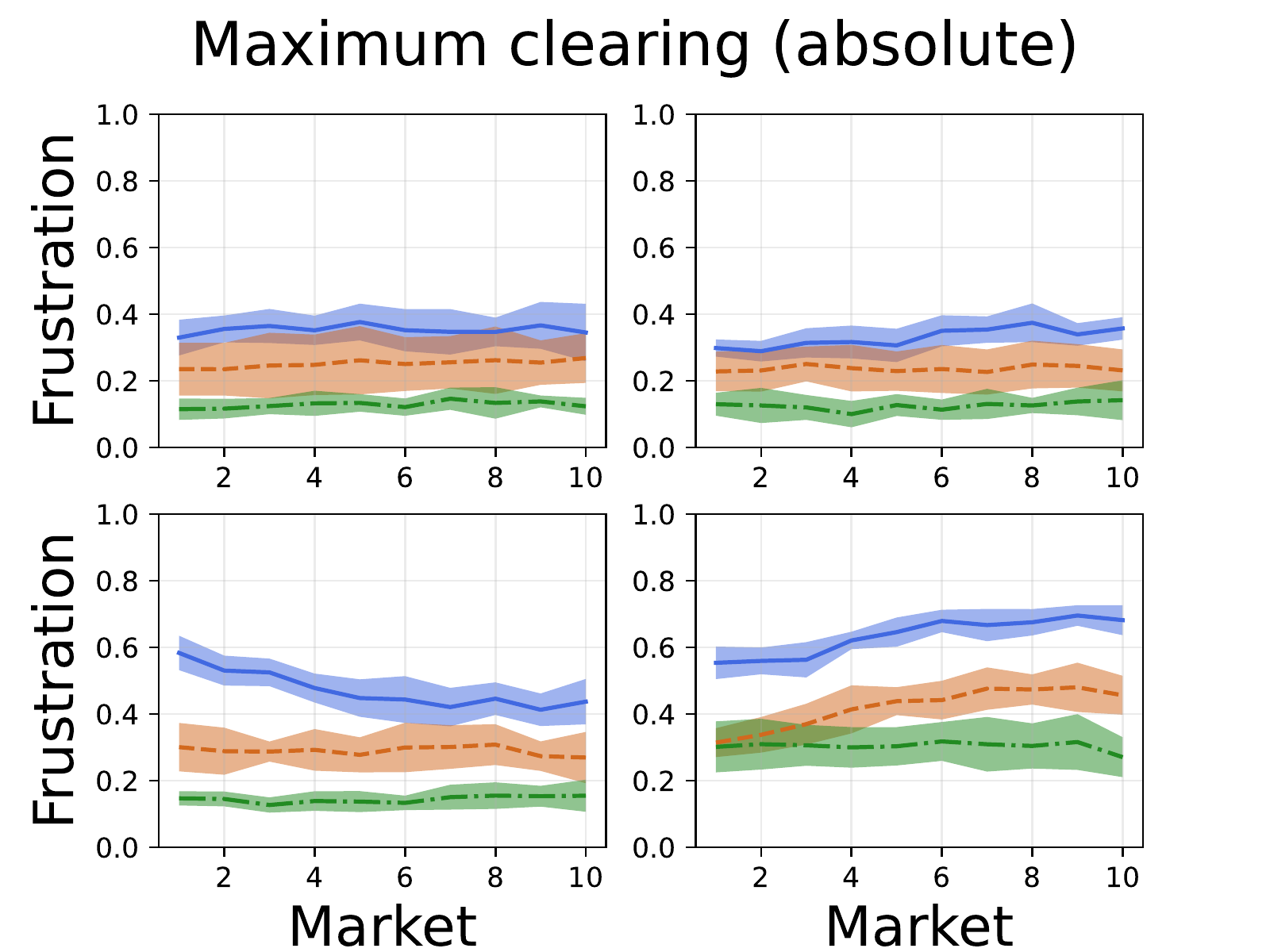} 
\includegraphics[width=.24\textwidth]{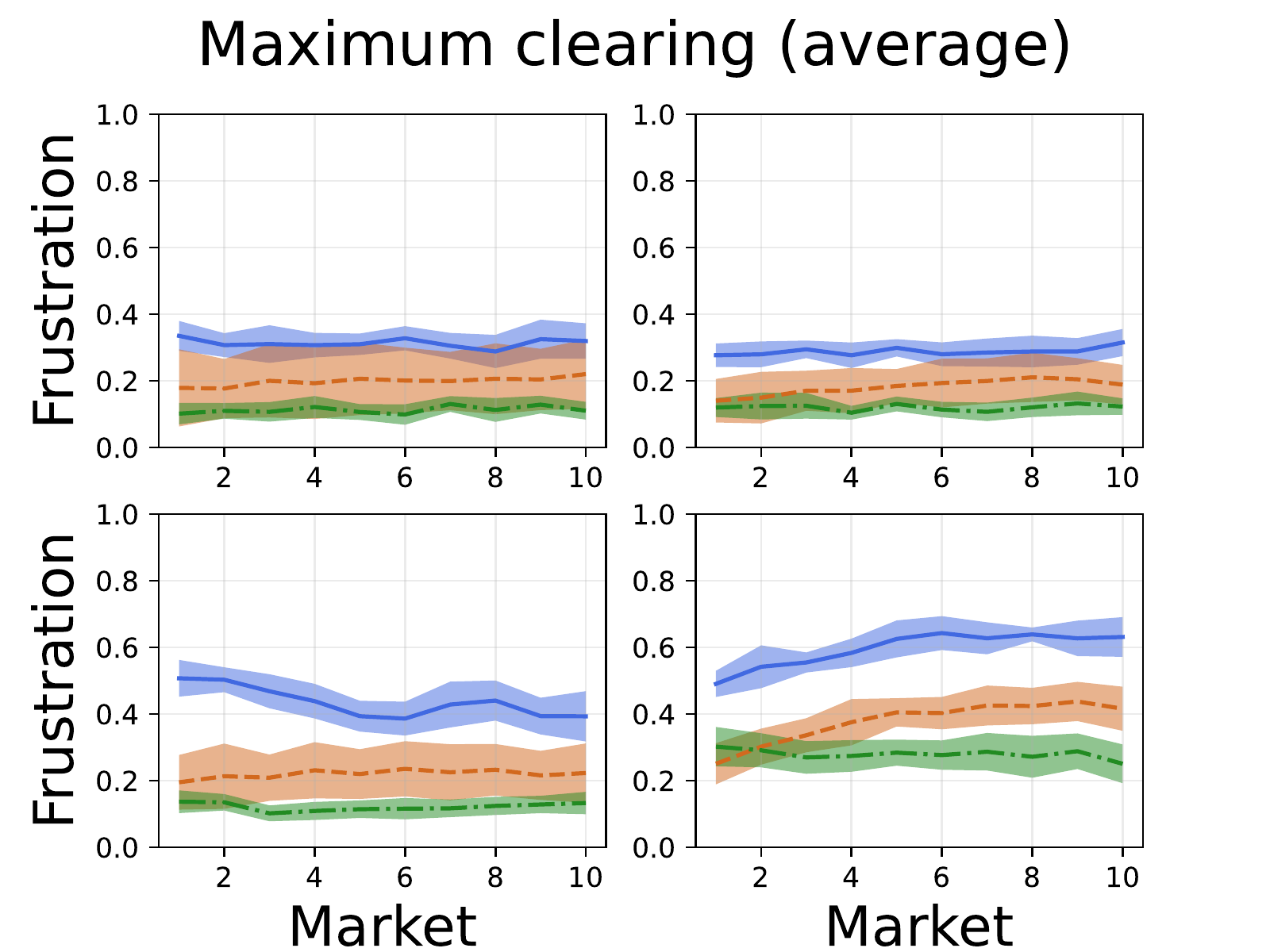} 
\caption{(Top) The Price of Anarchy and (Bottom) the individual frustrations of the four buyers in systems with three variants of fairness for four different market mechanisms. The poor buyer is always bottom right in the frustration graphs.} 
\label{fig:poa}
\end{figure*}

\subsection{Scalability}\label{ssec: scalability}
In the second experiment we explore the computational time as a function of the number of participants. More specifically, we study Crises with $N$ buyers and $N$ sellers. In these simulations, we generate demands and earning as described earlier, but for more traders. Each buyer has a chance to be rich with probability 3/4 and poor otherwise. The computational time required for each $N$ is depicted 
in Figure \ref{fig: time complexity} on the left, for a system with the greedy market mechanism and fairness mechanism present, using $k=1$. For each value of $N$ we sampled 10 instances. The algorithm exhibits near-linear time complexity in the number of buyers and sellers $N$.

To show that we reached an approximate equilibrium even with a large number of traders, we again evaluate the NashConv. However, computing the NashConv takes a significant portion of the computational time. To investigate purely the scalability of the algorithm
we evaluate it once at the end of the training. The NashConv values, normalized by the number of buyers to facilitate comparison, are shown in Figure \ref{fig: time complexity} on the right. The results suggest we reached close approximations of the equilibria.

\subsection{Price of Anarchy over Markets}\label{subsec: robustness}

Here we present our main results: the empirical study of how the Price of Anarchy evolves in our hybrid system, in comparison to an intervention-free market. The results are depicted in Figure~\ref{fig:poa}, and they show the prices the society pays for distributing the critical Goods through a (regulated) market instead of centrally. All results are averaged over 10 instances and show also the standard errors. The top row compares the Prices of Anarchy of systems with the three earlier described modes of fairness for the four introduced Market mechanisms. Note that the PoA is always lower in the systems with Rights. Moreover, introducing a second trading period further decreases it. Another noteworthy observation is that maximum clearing allocations offer lower PoA than the other two, more basic mechanisms. 
    
The bottom row then shows the individual frustrations of the buyers. As expected, the poor buyer experiences the highest frustration. Otherwise the results observed with overall PoA clearly translate into the frustration of each buyer as well. Interestingly, the results suggest that introducing the fairness mechanism into the trading is beneficial not only for the poor buyer but for the rich buyers as well.

\subsection{Price of Anarchy over buyers and sellers} 
\label{app: PoA number of players}

\begin{figure*}[t] 
\centering
\centering
\includegraphics[width=0.24\textwidth]{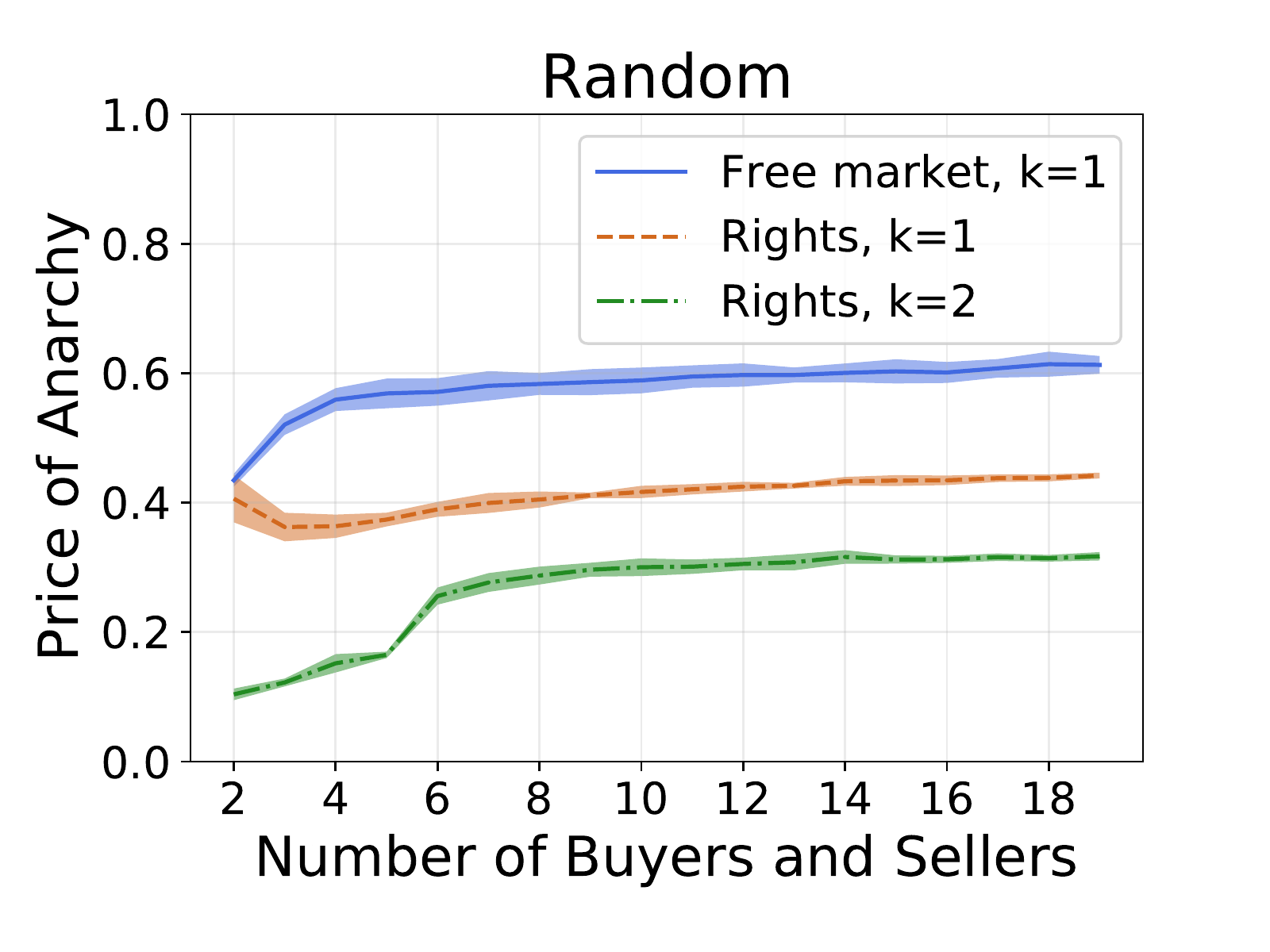}
\includegraphics[width=0.24\textwidth]{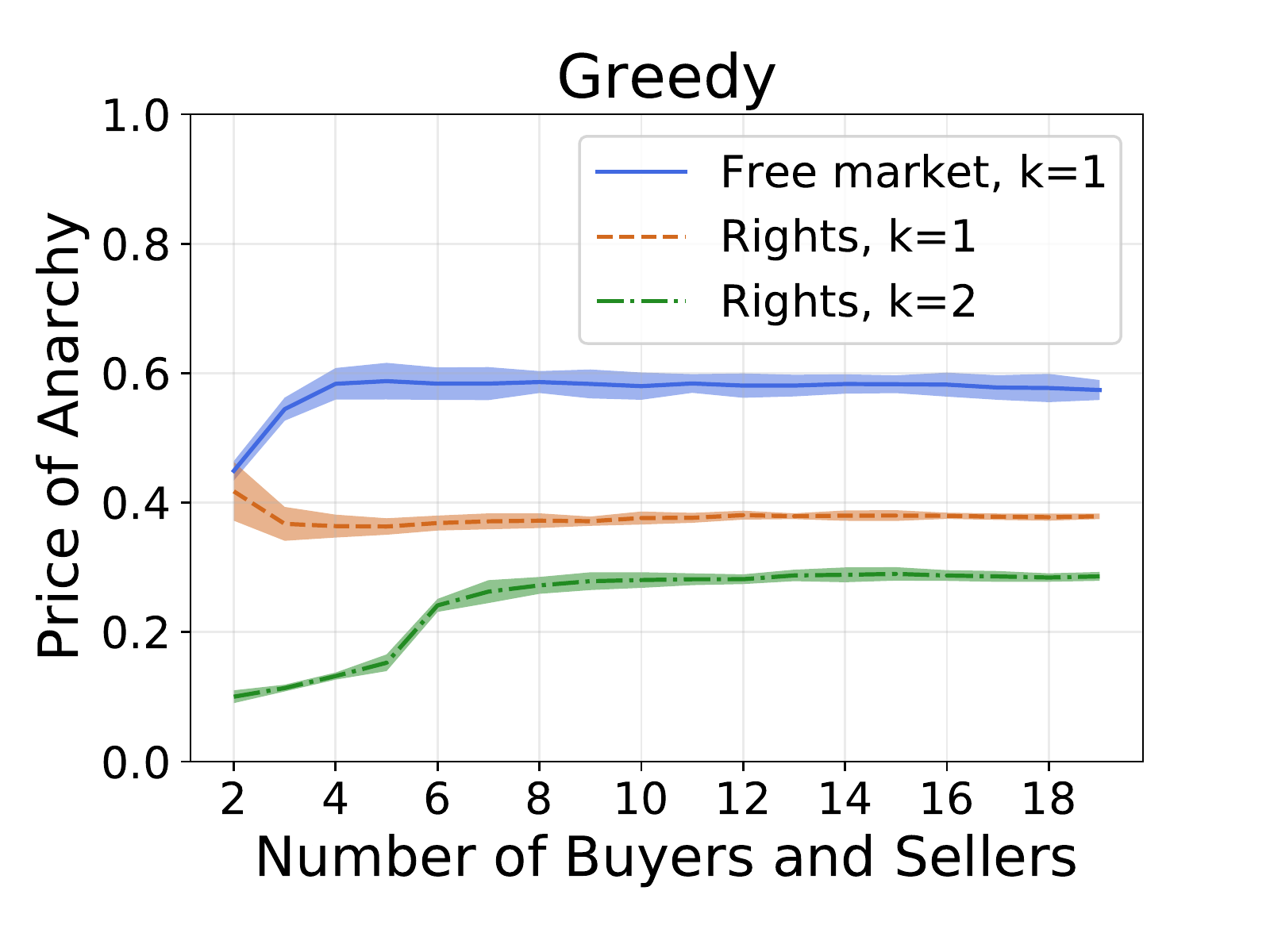}
\centering
\includegraphics[width=0.24\textwidth]{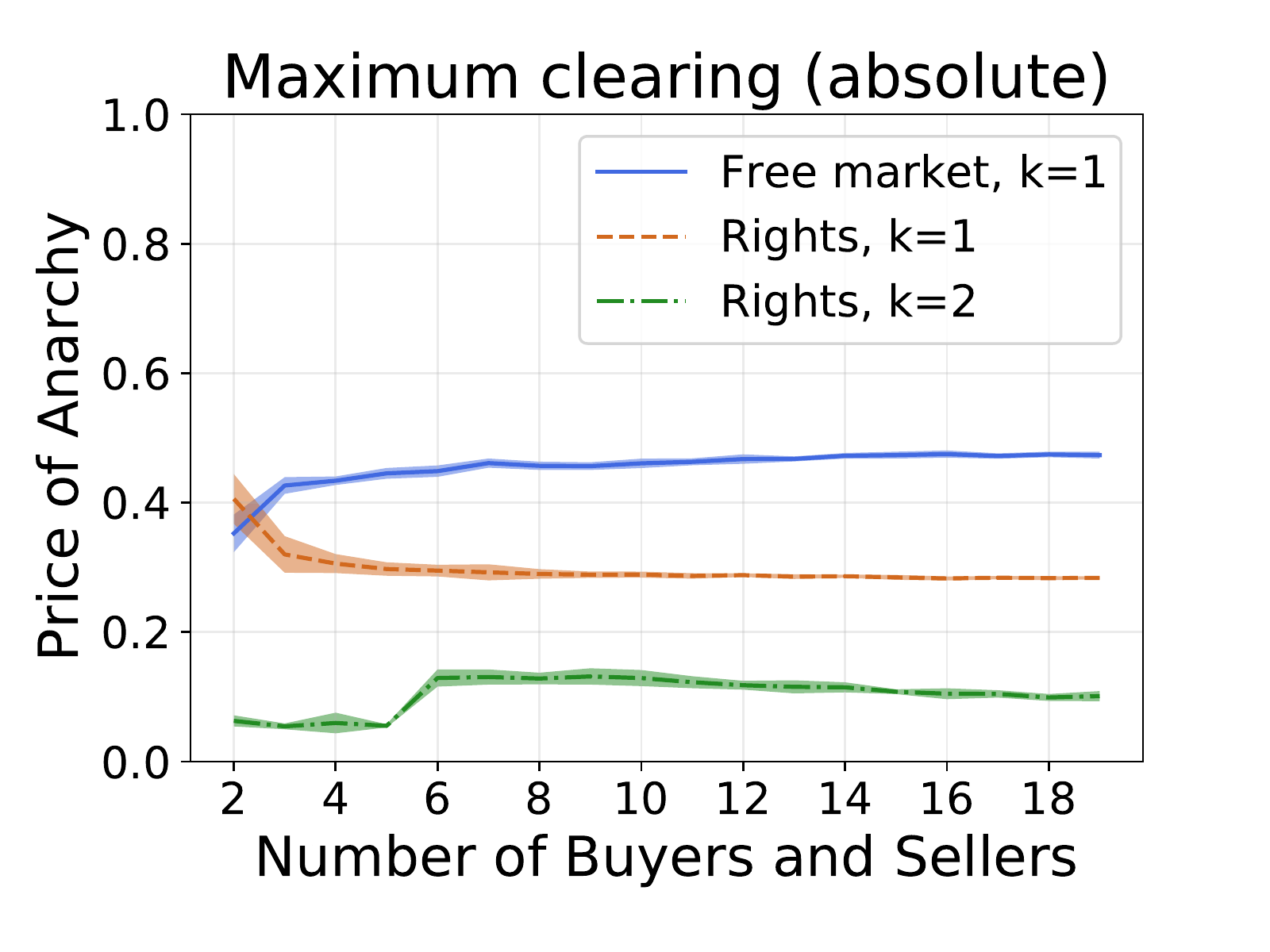}
\includegraphics[width=0.24\textwidth]{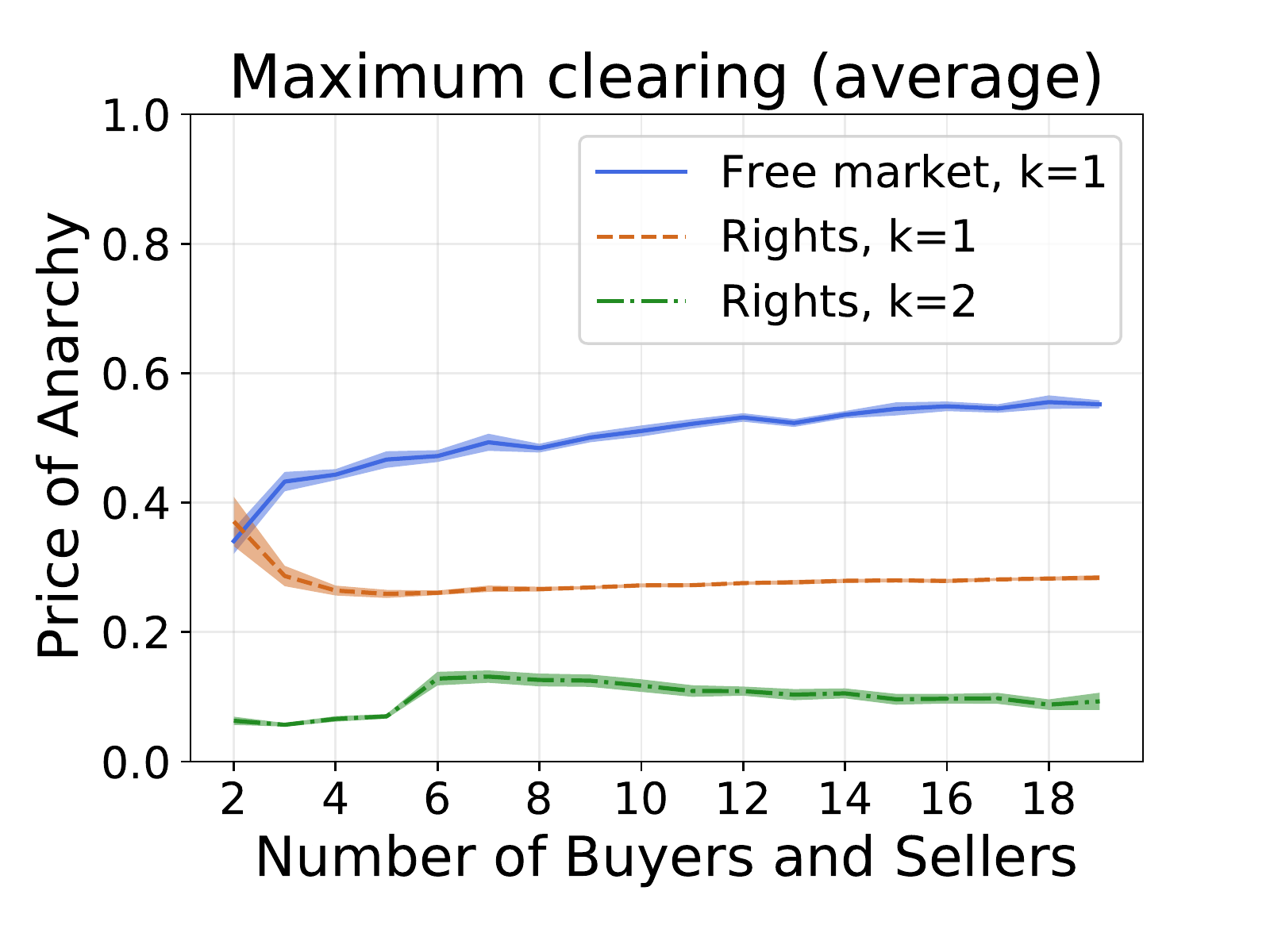}
\caption{The Price of Anarchy as a function of the number of buyers and sellers in systems with three variants of fairness for four different market mechanisms.} 
\label{fig:poa:traders}
\end{figure*}

We generate random instance of Crises with $N$ buyers and $N$ sellers for each $N\in\{2,\dots,19\}$ from the experimental domain described above.
The PoA as a function of $N$ in Figure~\ref{fig:poa:traders}. For each $N$ we generate 10 instances and the graphs show also the standard errors. The same general trend we reported above clearly translate over all the mechanism here as well: (i) \textit{the PoA is persistently the largest in the free market}, as opposed to the markets with Rights, and (ii) the market with $k=2$ buyers' stages dominates the market with $k=1$ stage. However, the results also indicate that different market mechanisms perform vastly differently. With the \textbf{random} mechanism, the PoA seem to slowly yet gradually increase with the number of traders, while in the \textbf{greedy} mechanism the PoA reaches a constant. The most interesting cases are the \textbf{maximum clearing} mechanisms. With both absolute and average prices, the PoA in the free market keeps increasing, while the PoA in the system with Rights
and $k=1$ iteration of the buyers' stage mostly stagnates. In the redistribution system with $k=2$ buyers' stages, the PoA after $N=6$ 
consistently decreases, reaching a PoA value $\approx 5.93$-times smaller than in the free market.

\section{Conclusion}

To the best of our knowledge, we are the first to introduce a system explicitly combining a \textit{double-sided} market mechanism with a fairness mechanism allocating buying rights for more socially just redistribution of critical goods during the times of need. We adopted the contested garment distribution as a baseline method for fair allocation of rights and studied four separate market mechanism for trading: random, greedy, absolute-prices maximum clearing, and average-prices maximum clearing. Our two main theoretical results show that the last two allocations can be computed in polynomial time. We then defined an analogue of Price of Anarchy (PoA) in our system as a sum of so-called individual frustrations, which are scaled differences between the amount of goods each trader was entitled to according to the fairness mechanism and the amount they were actually able to secure in the market. Furthermore, we developed a reinforcement-learning algorithm capable of approximating an equilibrium of the system in order to evaluate the PoA in practice. In the last part of our work, we show on a notorious example of a system with an underfunded and short-supplied buyer that introducing the buying rights may significantly decrease the frustrations, ergo, the PoA, especially for mechanisms prioritizing the amount of goods sold. Yet, it still remains an open question whether there exists a mechanism admitting zero PoA in the limit.

\paragraph{Future work}

We see two major limitations of our work. First, we focused on the full-blown crises and assumed a small, but constant resupply of goods and money over many trading periods. We would like to study the consequences of more complex system dynamics akin to, e.g., the bullwhip effect caused by a steep increase in demand at the beginning of a crisis. Second, we restricted our fairness model to the contested garment distribution. Considering other models may change the system dynamics, and perhaps improve the PoA. Moreover, designing an optimal fairness rule may be done similarly as for voting mechanisms in the manner of~\cite{Koster2022}.

\newpage

\section*{Acknowledgments}
This research was supported by the CRISDIS project of the Czech Ministry of the Interior no.VI04000107 and SIMTech-NTU Joint Laboratory on Complex Systems. Computational resources were supplied by the project \enquote{e-Infrastruktura CZ} (e-INFRA CZ LM2018140) supported by the Ministry of Education, Youth and Sports of the Czech Republic. The authors would also like to express their gratitude towards an anonymous reviewer who suggested to investigate the effects of the number of traders on the Price of Anarchy.

\balance

\bibliographystyle{ACM-Reference-Format} 
\bibliography{main}

\clearpage

\appendix
\section{Hyperparameters} \label{app:hyp}
The experiments used the following values of parameters:

\medskip

\noindent\begin{tabular}{ll}
Actor learning rate & $3\cdot 10^{-4}$ \\
Critic learning rate & $ 10^{-3}$ \\
Actor hidden layer size & 32 \\
Critic hidden layer size & 256 \\
Batch size & 512 \\
L2 penalty & $10^{-2}$ \\
Discount factor $\gamma$ & $0.99$ \\
Target network update rate & $0.002$ \\
Actor training frequency & 3 \\
Entropy penalty & $3\cdot 10^{-3}$ \\
Training episodes & 3000 \\
NashConv training episodes & 100
\end{tabular}

\section{Example of a simple redistribution system}
\label{app: toy example}
Consider an example of a redistribution system with two buyers $\{b,b'\}$ and a single seller $s$. Let the buyer $b$ have more funds and the other buyer $b'$ have a larger demand, formalized as $({\cal M}^1_b, {\cal M}^1_{b'}) = (2,1)$ and $({\cal D}^1_b, {\cal D}^1_{b'}) = (1,3)$. In addition, let the seller have two units of Good for sale. 

Assume the seller offers their entire stock ($v^G = 2$) for sale at price $p^G=1$, resulting in a distribution of Rights $({\cal R}^1_b, {\cal R}^1_{b'}) = (0.5,1.5)$ as dictated by the contested garment rule. After releasing the Rights to the buyers, the second buyer $b'$ realizes they may directly buy a unit of Good with the unit of Money they have and puts the remaining half a unit of Right for sale. Assume they choose a price $p^G_{b'} = 1$. The other buyer $b$ has no intention to offer their Right for sale as they prefer to use it for buying the Good and $b'$ is unlikely to buy the Right anyway. Let the acceptable prices and volumes be declared as
\begin{align}
    (v_b^R, p_b^R, \overline{v}_b^R, \overline{p}_b^R, \overline{v}_b^G, \overline{p}_b^G) &= 
    (0, 0, 0.5, 1, 1, 1),\\
    (v_{b'}^R, p_{b'}^R, \overline{v}_{b'}^R, \overline{p}_{b'}^R, \overline{v}_{b'}^G, \overline{p}_{b'}^G) &= 
    (0.5, 1, 0, 0, 1, 2).
\end{align}

Next, the bids are cleared using a market mechanism. In this example we will show the greedy allocation defined in Subsection 3.2 because it is the most transparent market mechanism we consider in the paper. 
\begin{itemize}
    \item We start with buyer $b'$ since $\overline{p}_{b'}^G > \overline{p}_{b}^G$: 
    \begin{enumerate}
        \item Because the acceptable price $\overline{p}_{b'}^G$ is greater than $p^G$, in the first stage the volume of $\text{min}\left\{v^G, \overline{v}_{b'}^G, R_{b'}-v_{b'}^R, \frac{{\cal M}^1_{b'}}{p^G}\right\}=1$ is sold to buyer $b'$. 
        \item In the second stage $b'$ buys nothing since their desired volume and desired price of Right are zero and they already bought $\overline{v}_{b'}^G$ of Good.
    \end{enumerate}
    \item We proceed with buyer $b$:
    \begin{enumerate}
        \item Since $\overline{p}_{b}^G \ge p^G$, buyer $b$ uses the Right they were allocated to buy $\text{min}\left\{v^G-1, \overline{v}_{b}^G, R_{b}-v_b^R, \frac{{\cal M}^1_{b}}{p^G}\right\}=0.5$ units of Good in the first stage.
        \item Finally, in their second stage $b$ buys the Right and the Good in equal quantities from $b'$ and $s$, respectively, buying 
        $$\text{min}\left\{v^G - 1.5, \overline{v}^R_b, \overline{v}_{b}^G - 0.5, \frac{{\cal M}^1_{b}-0.5 p^G}{p^G + p_{b'}^R}\right\}=0.5$$
        units of Good. The Money $b$ pays is split between $s$ and $b'$ proportionally to $p^G$ and $p^R_{b'}$, which in this case is half and half.
    \end{enumerate}
\end{itemize}
After the market mechanism reallocates the resources in this manner, the buyers end up with $(1, 1)$ units of Good and $(0.5, 0.5)$ units of Money. The seller also sells all the Good they had for trade.

The redistribution system ends the trading iteration by using the transition function to subtract the buyers' demands from their Good, simulating the consumption. The buyers hence have no Good left and they both receive utility one. The transition function also simulates the buyers' external earnings and sellers' resupply, providing them with extra Money and Good for the following Market. The model assumes the amounts of these periodic additional resources are held constant through the entire Crisis and are equal to the Money and Good the traders had at the beginning of the first Market. To sum up, this means that at the beginning of the second iteration of the redistribution system the buyers have $(2.5, 1.5)$ units of Money and the seller has two units of Good. 

\medskip

In contrast to free market, the redistribution system with Rights hence guarantees buyer $b'$ the extra 0.5 units of Money at the start of the second Market. This allows them to buy more Good in the future and decrease their accumulating frustration. How much the frustration decreases depends not only on the strategies of the traders but also on the market mechanism. If we employ the maximum clearing mechanisms, the result does not change in this simple case since there is no other way to clear the bids.

\section{Proofs}
\label{app: proofs}

\begin{theorem}
\label{thm.max}
Maximum clearing allocation can be found efficiently using a reduction to the Max Flow problem. As a consequence, a Maximum clearing allocation is polynomial for both divisible and indivisible Good.
\end{theorem}
\begin{proof}
 Given the disjoint copies of the graphs $G_G, G_R$, we construct an instance of the Max Flow problem as follows:
\begin{enumerate}[leftmargin=.3cm]
\item introduce two new vertices $s,t$;

\item join $s$ by an arc $(s,v)$ to each vertex $v$ of $\mathcal{B}$ in $G_G$. Let the capacity $cap(s,v)$ of this arc be equal to the amount of the {\em remaining rights} of $v$, i.e., the assigned amount minus the amount intended to be sold. Clearly, each buyer desires to buy at least $cap(s,v)$ of Good.

\item  join $s$ by an arc $(s,v)$ to each vertex $v$ of $\mathcal{B}_S$ in $G_R$. Let the capacity of this arc be equal to $w_R(v)$, i.e., the amount (possibly zero) of Right $v$ intends to sell;

\item  orient each edge of $G_R$ towards $\mathcal{B}_B$, the capacity of $(x,y)$ being equal to $w_R(y)$, i.e., the amount of Right $y$ intends to buy;

\item  orient each edge of $G_G$ towards $\mathcal{S}$, the capacity of $(x,y)$ being equal to $w_G(y)$, i.e., the amount of Good $y$ intends to sell;
 
\item introduce a copy $\mathcal{B}'$ of $\mathcal{B}$ and join each vertex $v\in \mathcal{B}_B$ of $G_R$ by an arc $(v,v')$ to its copy $v'\in \mathcal{B}'$, its capacity being $w_R(v)$, i.e., the amount of Right $y$ intends to buy;

\item  join each $v'\in \mathcal{B}'$ to $\mathcal{S}$ in the same way as its copy $v$ is joined to $\mathcal{S}$ in $G_G$, orient these new edges towards $\mathcal{S}$ and let the capacity of each such arc terminating in $y\in \mathcal{S}$ be $w_G(y)$, i.e., the amount of Good $y$ intends to sell;

\item  join each vertex $y$ of $\mathcal{S}$ to $t$ by the arc $(y,t)$, its capacity being $w_G(y)$, i.e., the amount of Good $y$ intends to sell.
\end{enumerate}


This finishes the construction of the instance of the Max Flow problem. It is straightforward to see that max flow from $s$ to $t$ provides a clearing of bids with the maximum amount of the Good sold. Also, it is ensured that Right is bought along with the same amount of Goods.
\end{proof}

\begin{theorem}
\label{thm.max2}
Maximum clearing allocation with average bids can be found efficiently using a linear program.
\end{theorem}
\begin{proof}
We can find the maximum clearing allocation using the following linear program, where the variable $r_{b,b'}$ represents the amount of Right sold to $b'$ by $b$, with $(b,b')\in E_R$ and the variable $g_{s,b}$ represents the amount of Good sold to $b$ by $s$, with $(s,b)\in E_G$. We also introduce the variables $m$ and $M$, representing the minimal, resp maximal, amount of Good bought by a buyer. Furthermore, we define $c$ as $c=\epsilon*U$ where $\epsilon$ is the desired sensibility of the objective function and $U$ an upper bound on $(M-m)$: $U = \max\limits_{b\in B} \left(\min(d_b, \sum\limits_{(s,b)\in E_G} w_G(s))\right)$; $c$ will be used to normalize $(M-m)$ in order to not interfere with the rest of the objective function. In our experiments, we used $c=\frac{1}{1000}$.
{\small
\begin{equation*}
\begin{array}{ll@{}lll}
    \max \sum\limits_{(s,b)\in E_G} g_{s,b}  - c( M-m)&& (1)\\
    
    \text{s.t.} \\
    
    \sum\limits_{(s,b)\in E_G}g_{s,b}\leq r_b+\sum\limits_{(b',b)\in E_R}r_{b',b}-v^R_b~ & \forall b\in B&(2) \\
    
    \sum\limits_{(s,b)\in E_G}g_{s,b}\leq \overline v^G_b~ & \forall b\in B&(3) \\
    
    \sum\limits_{(s,b)\in E_G}g_{s,b}\leq v^G_s~ & \forall s\in S&(4) \\
    
    \sum\limits_{(b,b')\in E_R}r_{b,b'}\leq v^R_b~ & \forall b\in B&(5) \\
    
    m \leq \sum\limits_{s\in S} g_{s,b} ~ & \forall b\in B&(6) \\
    
    M \geq \sum\limits_{s\in S} g_{s,b} ~ & \forall b\in B&(7) \\
    
    \sum\limits_{(s,b)\in E_G}g_{s,b}*p^G_s \leq \overline p^G_b\sum\limits_{(s,b)\in E_G}g_{s,b}~ & \forall b\in B&(8) \\
    
    \sum\limits_{(b',b)\in E_R}p^R_{b'}*r_{b',b} \leq \overline p^R_b\sum\limits_{(b',b)\in E_R}r_{b',b}~ & \forall b\in B&(9) \\
    
    \sum\limits_{(s,b)\in E_G}p^G_s g_{s,b} + \sum\limits_{(b',b)\in E_R}p^R_{b'} r_{b',b}\leq M_b & \forall b\in B&(10)\\
    g_{s,b}\geq 0~ & \forall (s,b)\in E_G~&(12)\\
    r_{t,b}\geq 0~ & \forall (t,b)\in E_R~&(13)\\
\end{array}
\end{equation*}
}%

In this linear program, the objective function (1) maximizes the exchanges of goods, and spreads the distribution over the buyers. The constraints (2) and (3) then enforce that the buyers buy less Good than they have Rights, and the amount of Good they buy does not exceed their desired volume $v^G_b$. The constraint (4) imposes a restriction on the amount of Good the sellers may sell, ensuring it is at most $v_s^G$, i.e., the amount they committed themselves to be willing to sell. Similarly, the constraint (5) imposes that the buyers selling Good sell at most the amount they intend to sell $v^R_b$. The constraint (6), resp (7), assures that m is lower, resp. higher, than the minimal, resp. maximal, amount of good bought by a buyer, and the sense of the objective function ensure that it will be exactly this quantity. The constraint (8) imposes that the buyers pay at most in average $p^G_b$ for the Good. The constraint (9) forces that the buyers pay at most in average $p^R_b$ for the Right. The constraint (10) is the budget constraint. 
\end{proof}

\end{document}